\documentclass{ptephy}
\preprintnumber{XXXX-XXXX}

\usepackage{boites}
\numberwithin{equation}{section}
\usepackage{amsmath}	

\makeatletter
\DeclareRobustCommand{\cev}[1]{%
  \mathpalette\do@cev{#1}%
}
\newcommand{\do@cev}[2]{%
  \fix@cev{#1}{+}%
  \reflectbox{$\m@th#1\vec{\reflectbox{$\fix@cev{#1}{-}\m@th#1#2\fix@cev{#1}{+}$}}$}%
  \fix@cev{#1}{-}%
}
\newcommand{\fix@cev}[2]{%
  \ifx#1\displaystyle
    \mkern#23mu
  \else
    \ifx#1\textstyle
      \mkern#23mu
    \else
      \ifx#1\scriptstyle
        \mkern#22mu
     r \else
        \mkern#22mu
      \fi
    \fi
  \fi
}

\begin{document}

\title{Gauge symmetry breaking in the adiabatic self-consistent
collective coordinate method}

\author{\name{\fname{Koichi} \surname{Sato}}{1} 
}

\address{\affil{1}{Department of Physics, Osaka City University, Osaka 558-8585, Japan}
\email{satok@sci.osaka-cu.ac.jp}}

\begin{abstract}%
We study gauge symmetry breaking by adiabatic approximation 
in the adiabatic self-consistent collective coordinate (ASCC) method.
In the previous study, we found that the gauge symmetry of the equation
 of collective submanifold is (partially)
broken by its decomposition into the three moving-frame equations depending on the order of $p$.
In this study, we discuss the gauge symmetry breaking 
by the truncation of the adiabatic expansion.
A particular emphasis is placed on the symmetry under the gauge
transformations which are not point transformations.
We also discuss a possible version of the ASCC method 
including the higher-order operators which can keep the gauge symmetry.

\end{abstract}

\subjectindex{xxxx, xxx}

\maketitle


\section{Introduction}\label{sec1}

The adiabatic self-consistent collective coordinate (ASCC) method~\cite{Matsuo2000}
is a practical method for describing large-amplitude collective motion
of atomic nuclei with superfluidity~\cite{Matsuyanagi2010,Klein1991}.
It is an adiabatic approximation to the self-consistent collective
coordinate (SCC) method~\cite{Marumori1980, Matsuo1986a} and can be regarded as
an advanced version of the adiabatic time-dependent
Hartree-Fock-Bogoliubov (ATDHFB)
theory. The ASCC method overcomes the difficulties several versions of the
ATDHF(B) theory encountered (see \cite{Klein1991} for a review). 
It also provides with a non-perturbative scheme to solve the
basic equations of the SCC method and is applicable to
large-amplitude collective dynamics 
which is not accessible with the $(\eta, \eta^*)$ expansion method of the SCC method.

The ``gauge'' symmetry in the ASCC method was first pointed out by
Hinohara et al~\cite{Hinohara2007}.
They encountered a numerical instability in solving the basic equations
of the ASCC method, the moving-frame HFB \& QRPA equations.
They found that the instability was caused by symmetry under some
continuous transformation, under which the basic ASCC equations are
invariant. Because the transformation changes the phase of the state
vector, it is called the gauge symmetry. They proposed a prescription
to fix the gauge and successfully applied it to multi-O(4) model \cite{Hinohara2007} and
to the shape coexistence/mixing in proton-rich Se and Kr isotopes with
the pairing-plus-quadrupole model~\cite{Hinohara2008, Hinohara2009}.

After the successful application of the one-dimensional ASCC method,
an approximated version of the two-dimensional ASCC method, the constrained
HFB plus local QRPA method, was proposed \cite{Hinohara2010b} 
and applied to large-amplitude quadrupole collective dynamics 
\cite{Sato2011,Watanabe2011,Hinohara2011a,Hinohara2011,Hinohara2012,Yoshida2011, Sato2012}.
(We mean by the $D$-dimensional ASCC method that the dimension 
of the collective coordinate $q$ is $D$.)
However, little progress had been made in the understanding on the gauge
symmetry in the ASCC method.

Quite recently, we investigated the gauge symmetry in the ASCC method
on the basis of the Dirac-Bergmann theory of constrained
systems, which brought about a new insight~\cite{Sato2015}. 
According to the theory of constrained systems initiated by Dirac and
Bergmann
~\cite{Dirac1950,Anderson1951,Dirac1964},
the gauge symmetry is associated with constraints 
which are originated from the singularity of the Lagrangian.
In the ASCC method, the linear term of $n$ in the collective Hamiltonian
plays the role of a constraint and leads to the gauge symmetry.
In Ref. \cite{Sato2015}, we discussed possible gauge transformations 
in the ASCC method from a general point of view based on the Dirac-Bergmann theory of
constrained systems.
We found that the four examples or four types of the gauge
transformations play essential roles 
to discuss the gauge symmetry and its breaking 
and to determine the form of the general gauge transformation
under which the ASCC equations are invariant.
[The four examples are listed in Eqs. (\ref{eq:^q' ex1})-(\ref{eq: N' in ex 4}).]

The basic equations of the ASCC method, i.e., 
the moving-frame HFB \& QRPA equations and the canonical-variable
conditions, are derived from the 
equation of collective submanifold and the canonicity conditions,
respectively, with use of the adiabatic expansion.
Although the equation of collective submanifold and the canonicity conditions are
invariant under the general gauge transformation,
the gauge symmetry of the moving-frame QRPA equations is partially broken.
For example, we found in Ref.~\cite{Sato2015} that
the gauge symmetry in the moving-frame QRPA equations is broken 
by the decomposition of the equation of collective submanifold into the three
moving-frame equations in Example 3. 
On the other hand, in Example 1, 
the gauge symmetry of the moving-frame QRPA equations is broken 
unless $\hat Q$ and $\tilde N$ commute, but
it has not been elucidated why the gauge symmetry is broken in the moving-frame 
QRPA equations.
Besides the moving-frame QRPA equations,
the gauge symmetry of the canonical-variable conditions 
also can be regarded to be broken by the truncation of the adiabatic expansion,
as we show in this paper.

As mentioned above, the ASCC method is an adiabatic approximation to the
SCC method. The term ``adiabatic approximation'' is often used for different meanings.
In Ref. \cite{Matsuyanagi2016}, it is used for the approximate solution
of the equation of collective submanifold by taking into account up to
the second order in the adiabatic expansion with respect to the
collective momenta $p$.
In this paper, we shall use this term in a broader sense to 
mean an approximate solution of the equation of collective submanifold with use of the
adiabatic expansion up to a certain order with respect to $p$.
Precisely speaking, the ``approximate solution'' implies the following two things.
One is that the equation of collective submanifold is decomposed into 
a certain number of moving-frame equations depending on the order of $p$.
(In the ASCC method in this paper, we consider three moving-frame equations.)
The other is that the adiabatic expansion is truncated up to a certain order.
These ``decomposition'' and ``truncation'' are sources of the gauge
symmetry breaking in the ASCC method.

In this paper, we investigate the gauge symmetry breaking 
by the adiabatic approximation   
in the ASCC method and discuss a possible
extension of the ASCC method to keep the gauge symmetry by including the higher-order terms.
In the ASCC method, we assume the following form of the state vector
\begin{align}
 |\phi (q,p,\varphi, n)\rangle = e^{-i\varphi \tilde N}|
 \phi(q,p,n)\rangle 
= e^{-i\varphi \tilde N} e^{i\hat G(q,p,n)}| \phi(q)\rangle. \notag  
\end{align}
Here, $q$ is a collective coordinate and $p$ is its conjugate momentum.
$\varphi$ is the gauge angle conjugate to the particle number $n=N-N_0$.
$\tilde N=\hat N-N_0$ is the particle number operator measured from a
reference value $N_0$.
In the ASCC method we have considered so far, $\hat G$ is expanded up to
the first order with respect to $(p,n)$.
We show that the gauge symmetry breaking of the moving-frame QRPA
equations in Example 1 is due to the truncation of the expansion of
$\hat G$ to the first order.

The paper is organized as follows. 
In Sect. 2, we discuss the gauge symmetry of the canonicity conditions
and that of the canonical-variable conditions. We show that the gauge symmetry
of the canonical-variable conditions under the non-point gauge transformations 
is broken by 
the truncation of the adiabatic expansion.
In Sect. 3, we expand $\hat G$ up to the second order
and show that the gauge symmetry breaking at the order of $p$ is
recovered if the second-order operators in $\hat G$ are taken into
account.
We illustrate how the higher-order operators are transformed under the gauge
transformations.
In Sect. 4, a version of the ASCC method including the third-order
operators is given. With the third-order operators, 
the moving-frame HFB~\&~QRPA equations and the canonical-variable
conditions  up to the second order
are gauge invariant under the
non-point gauge transformations.
The concluding remarks are given in Sect. 5.
In Appendix A, the derivation of the basic equations 
for the second-order and third-order expansions of $\hat G$
is given.
In Appendix B, we present the gauge transformations of the third-order operators in Examples 2-4.

As in Ref. \cite{Sato2015}, we consider the one-dimensional ASCC method with a single component 
for simplicity in this paper.
However, the extension to the multi-dimensional and multi-component cases is straightforward.

\section{Canonicity conditions and gauge transformation}

\subsection{Canonicity and canonical-variable conditions}
We consider gauge transformations of the canonicity conditions
and those of the canonical-variable conditions, which are derived from 
the canonicity conditions with use of the adiabatic expansion.
As in Ref. \cite{Sato2015}, we assume the state vector in the following
form.
\begin{align}
 | \phi(q,p,\varphi, n) \rangle &= e^{-i \varphi \tilde N
  | }|\phi(q,p,n)\rangle \label{eq:hin2.3},\\  
 | \phi(q,p, n) \rangle &= e^{i \hat G }|\phi(q)\rangle , \label{eq:phi(q,p,n)}\\
\hat G(q,p,n)&=p\hat Q(q) +n\hat \Theta (q).
\end{align}
The canonicity conditions are given by
\begin{align}
\langle \phi(q,p,\varphi,n) | \mathring{Q}| \phi (q,p,\varphi,n) \rangle
=&\langle \phi(q,p,\varphi,n) | \frac{1}{i}\partial_p | \phi (q,p,\varphi,n) \rangle
=-\frac{\partial s}{\partial p}, \label{eq:canonicity 1}\\  
\langle \phi(q,p,\varphi,n) | \mathring{P}| \phi (q,p,\varphi,n) \rangle
=&\langle \phi(q,p,\varphi,n) | i\partial_q| \phi (q,p,\varphi,n) \rangle
 =p+\frac{\partial s}{\partial q}, \label{eq:canonicity 2}\\  
\langle \phi(q,p,\varphi,n) | \mathring{\Theta}| \phi (q,p,\varphi,n)
 \rangle
=&\langle \phi(q,p,\varphi,n) | \frac{1}{i}\partial_n| \phi (q,p,\varphi,n) \rangle
 =-\frac{\partial s}{\partial n}, \label{eq:canonicity 3}\\  
\langle \phi(q,p,\varphi,n) | \tilde N| \phi (q,p,\varphi,n) \rangle
=&\langle \phi(q,p,\varphi,n) | i\partial_\varphi| \phi (q,p,\varphi,n) \rangle
=n+\frac{\partial s}{\partial \varphi}.\label{eq:canonicity 4}
\end{align}
Here, $s$ is an arbitrary function of $(q,p,\varphi, n)$ and is set to
$s=0$ in the ASCC method. As discussed in Ref. \cite{Sato2015},
$s$ is related to the generating function of a canonical transformation.

As shown in Ref. \cite{Matsuo2000}, 
$\mathring{P}^\prime =e^{-i\hat G} e^{i\varphi \tilde N}\mathring{P}e^{-i\varphi \tilde N} e^{i\hat G}$,
$\mathring{Q}^\prime =e^{-i\hat G}e^{i\varphi \tilde N} \mathring{Q}e^{-i\varphi \tilde N} e^{i\hat G}$,
$\mathring{\Theta}^\prime =e^{-i\hat G}e^{i\varphi \tilde N} \mathring{\Theta} e^{-i\varphi \tilde N}e^{i\hat G}$,
and $\tilde N^\prime =e^{-i\hat G}e^{i\varphi \tilde N} {\tilde N}e^{-i\varphi \tilde N} e^{i\hat G}$
are expanded as follows.
\begin{align}
\mathring{P}^\prime&=i\partial_q - p\partial_q \hat Q-n\partial_q \hat
 \Theta +\cdots,
\label{eq:ring P' 1st-order}\\ 
\mathring{Q}^\prime
&=\hat Q +\frac{i}{2}[\hat Q, p\hat Q+n\hat \Theta ] +\cdots
=\hat Q +n\frac{i}{2}[\hat Q, \hat \Theta ] +\cdots,\\
\mathring{\Theta}^\prime
&=\hat \Theta +\frac{i}{2}[\hat\Theta,p\hat Q+n\hat \Theta]+\cdots
=\hat \Theta +p\frac{i}{2}[\hat\Theta,\hat Q]+\cdots, \label{eq: ring
 Theta' 1st-order } \\
\tilde{N}^\prime&=\tilde N +  ip[ \tilde N, \hat Q] +in [ \tilde N, \hat
 \Theta] +\cdots .\label{eq:tilde N' 1st-order}
\end{align}
By substituting (\ref{eq:ring P' 1st-order})-(\ref{eq:tilde N' 1st-order}) into the canonicity conditions (\ref{eq:canonicity
1})-(\ref{eq:canonicity 4}) 
and setting $s=0$, we obtain the zeroth- and first-order canonical-variable conditions.

\noindent
\underline{Canonical-variable conditions}
\begin{align}
\langle \phi(q)|\,\hat Q   \,|\phi(q)\rangle =0, \label{eq:O(1) can.var.cond 1}\\
\langle \phi(q)|\,\hat P   \,|\phi(q)\rangle =0, \label{eq:O(1) can.var.cond 2}\\
\langle \phi(q)|\,\hat \Theta   \,|\phi(q)\rangle =0, \label{eq:O(1) can.var.cond 3}\\
\langle \phi(q)|\,\tilde N   \,|\phi(q)\rangle =0, \label{eq:O(1) can.var.cond 4}
\end{align}
\begin{align}
&\langle \phi(q)|\,[\hat Q, \hat P] \,|\phi(q)\rangle =i, \label{eq:can.var.cond 1} \\
&\langle \phi(q)|\,[\hat P, \hat \Theta] \,|\phi(q)\rangle=0, \label{eq:can.var.cond 2} \\
&\frac{i}{2}\langle \phi(q)|\,[\hat Q, \hat \Theta]\,|\phi(q)\rangle=0, \label{eq:can.var.cond 3}\\
&\langle \phi(q)|\,[\hat \Theta, \tilde N] \,|\phi(q)\rangle =i, \label{eq:can.var.cond 4} \\
&\langle \phi(q)|\,[\tilde N,\hat Q] \,|\phi(q)\rangle =0, \label{eq:can.var.cond 5}\\
&\langle \phi(q)|\,[\tilde N,\hat P] \,|\phi(q)\rangle =0, \label{eq:can.var.cond 6} 
\end{align}
The condition (\ref{eq:can.var.cond 6}) is obtained by differentiating
the condition (\ref{eq:O(1) can.var.cond 4}) with respect to $q$.
Here we have kept the factor $\frac{i}{2}$ in Eq. (\ref{eq:can.var.cond
3}), because it is necessary for discussion on the gauge symmetry later.

In Ref. \cite{Sato2015}, we found that the following four examples 
of the gauge transformations play essential roles for the discussion
on the gauge symmetry in the ASCC method.
Below we list the generators $G$ of the gauge transformations and the transformations
of collective variables and operators for each example.
(The generator $G$ should not be confused with $\hat G$ in
Eq. (\ref{eq:phi(q,p,n)}).)
Hereinafter let $\alpha$ be an infinitesimal.
Although a more general linear gauge transformation is given in
Ref. \cite{Sato2015}, it is not necessary for our purpose in this paper. 

\noindent
{Example 1:  $G = \alpha p n$ }
\begin{align}
&q \rightarrow q +\alpha n \label{eq:^q' ex1},\\
&\varphi \rightarrow \varphi +\alpha p \label{eq:^varphi' ex1}, 
\end{align}
\begin{align}
&\hat{Q} \rightarrow \hat{Q} +\alpha \tilde{N} \label{eq:^Q' ex1},\\
&\hat{\Theta} \rightarrow \hat{\Theta} +\alpha \hat{P} \label{eq:^Theta' ex1}.
\end{align}
{Example 2: $G = \alpha n^2/2$ }
\begin{align} 
&\varphi \rightarrow \varphi +\alpha n ,
\end{align}
\begin{align}
&\hat \Theta  \rightarrow \hat \Theta + \alpha \tilde N \label{eq:Theta' ex2}.
\end{align}
{Example 3: $G =  \epsilon n= \alpha qn$}
\begin{align} 
p      & \rightarrow  p   -\alpha n  \label{eq:p' ex3},\\
\varphi & \rightarrow \varphi + \alpha q,
\end{align}
\begin{align} 
\hat P      & \rightarrow  \hat P   -\alpha \tilde N  \label{eq:P' ex3},\\
\hat \Theta & \rightarrow \hat \Theta + \alpha \hat Q.
\end{align}
{Example 4:  $G = \epsilon n = \alpha \varphi n$}
\begin{align}
\varphi &\rightarrow \varphi +\alpha \varphi = e^\alpha \varphi, \\
n &\rightarrow n -\alpha n = e^{-\alpha} n,
\end{align}
\begin{align} 
\hat \Theta & \rightarrow  \hat \Theta + \alpha \hat \Theta =e^{\alpha}\hat \Theta \label{eq:Theta'  ex4},\\
\tilde N    & \rightarrow  \tilde N   -\alpha \tilde N = e^{ -\alpha}\tilde N \label{eq: N' in ex 4} .
\end{align}
As one can see from the canonicity conditions (\ref{eq:canonicity 1})-(\ref{eq:canonicity 4}),
$(q, \varphi)$ are coordinates and $(p,n)$ are the conjugate momenta.
Therefore, Examples 3 and 4 are point transformations.
Example 1 is the first example of the gauge transformations found 
in \cite{Hinohara2007}, 
and the general gauge transformation including Examples 1-4 is discussed in 
Ref. \cite{Sato2015}.

Before moving to the discussion on the canonicity conditions,
let us see the relation between the  transformations of the c-numbers
$(q,p,\varphi, n)$ and those of the corresponding operators.
Let us take Example 1.
In correspondence with (\ref{eq:^q' ex1})-(\ref{eq:^varphi' ex1}),
the differential operators are transformed as
\begin{align}
\partial_{p}&\rightarrow \partial_p-\alpha \partial_\varphi, \\ 
\partial_{n}&\rightarrow \partial_n-\alpha \partial_q ,
\end{align}
which leads to
\begin{align}
\mathring{Q} |\phi (q,p,\varphi, n)\rangle
\rightarrow (\mathring{Q} +\alpha \tilde{N}) |\phi (q,p,\varphi, n)\rangle, \label{eq:Q' ex1}\\
\mathring{\Theta} |\phi (q,p,\varphi,n)\rangle
\rightarrow (\mathring{\Theta} +\alpha \mathring{P})|\phi (q,p,\varphi,n)\rangle \label{eq:Theta' ex1}.
\end{align}
By considering the leading order with respect to $(p,n)$, we obtain
\begin{align}
\hat{Q} \rightarrow \hat{Q} +\alpha \tilde{N}, \tag{\ref{eq:^Q' ex1}}\\
\hat{\Theta} \rightarrow \hat{\Theta} +\alpha \hat{P}. \tag{\ref{eq:^Theta' ex1}}
\end{align}
When the transformation (\ref{eq:^Q' ex1})-(\ref{eq:^Theta' ex1}) is
applied to the state vector,
it implies
the transformation of the argument of the state vector as follows.
\begin{align}
 |\phi(q,p,\varphi,n) \rangle &\rightarrow e^{-i\varphi\tilde N}e^{ip(\hat Q(q) +\alpha\tilde N) +in(\Theta(q)+\alpha \hat P(q) )}  |\phi(q)\rangle \notag \\
&=e^{-i(\varphi      -\alpha p)\tilde N}e^{ip \hat Q(q-\alpha n)} 
e^{in\hat \Theta(q -\alpha n)} |\phi(q-\alpha n)\rangle \notag\\
&=|\phi(q-\alpha n, p, \varphi-\alpha p,n) \rangle 
,\label{eq:phi'} 
\end{align}
where we ignored the second-order terms with respect to $(p,n)$. Note that the signs of $\alpha$
in the last expression
are opposite to those in (\ref{eq:^q' ex1})-(\ref{eq:^varphi' ex1}).
Conversely, by transforming the argument of the state vector 
as $q\rightarrow q-\alpha n$, $\varphi \rightarrow \varphi-\alpha p$,
one can obtain the transformation (\ref{eq:^Q' ex1})-(\ref{eq:^Theta' ex1}).

\subsection{Gauge transformation of canonicity conditions}
In this subsection, we discuss the gauge symmetry of the canonicity conditions.
First,
we shall consider the gauge symmetry of the canonical-variable
conditions in Example 1.
One can easily ascertain that 
the canonical-variable conditions 
(\ref{eq:O(1) can.var.cond 1})-(\ref{eq:can.var.cond 6})
are invariant under the transformation of Example 1 (\ref{eq:^Q' ex1})-(\ref{eq:^Theta' ex1}).
For example, the condition
(\ref{eq:can.var.cond 3}) is transformed as follows.
\begin{align}
& \frac{i}{2}\langle \phi(q) | [\hat \Theta, \hat Q] | \phi (q) \rangle
 \notag \\
\rightarrow &\frac{i}{2}\langle \phi(q) | [\hat \Theta+\alpha \hat P, \hat Q+ \alpha
 \tilde N] | \phi (q) \rangle \notag \\
=&\frac{i}{2}\left(
\langle \phi(q) | [\hat \Theta, \hat Q] | \phi (q) \rangle 
+\alpha \langle \phi(q) | [\hat P, \hat Q] | \phi (q) \rangle
 \right. \notag \\
&\hspace{6em}\left. +\alpha \langle \phi(q) | [\hat \Theta, \tilde N] | \phi (q) \rangle 
+\alpha^2\langle \phi(q) | [\hat P, \tilde N] | \phi (q) \rangle 
\right)\notag \\
=&\frac{i}{2}\left(0- \alpha i+\alpha i+0\right)=0,
\end{align}
that is, the operators after the transformation
$\hat \Theta^\prime=\hat \Theta +\alpha \hat P, \hat Q^\prime=\hat Q+\alpha \tilde N$
also satisfy the weak canonical commutation relation
\begin{align}
  \frac{i}{2}\langle \phi(q) | [\hat \Theta^\prime, \hat Q^\prime] | \phi (q) \rangle=0.
\end{align}
In Refs. \cite{Hinohara2007} and \cite{Sato2015}, this
canonical-variable condition is said to be ``gauge invariant'' in this sense.
However, as we shall see below, this ``invariance'' of the canonical-variable
condition implies that the gauge symmetry is
actually broken.

In the following discussions, the arbitrary function $s$ in the
canonicity conditions plays a key role. 
When the set of canonical variables with the arbitrary function
$(q^i, p_i, s=0)$ is transformed to  $(q^{i \prime}, p_i^\prime, S)$
by a time-independent canonical transformation, 
the following relation holds~\cite{Yamamura1987}.
\begin{align}
 p_idq^i = p_i^\prime dq^{i\prime} +dS.
\end{align}
Thus, $S$ is the generating function of the time-independent canonical transformation. 
While $S=const.$ in Examples 3 and 4, which are point transformations,
in Examples 1 and 2, which are not point transformations, $S=-G$~\cite{Sato2015}.
Below we consider the gauge transformation of the canonicity conditions in
Examples 1 and 2.
In point transformations, $S$ does not contribute to
the canonicity conditions after the gauge transformation, because $S=const.$
On the other hand, in Examples 1 and 2, $S=-G$ does contribute as they are not point
transformations. 
We shall consider Examples 2 and 1. 

\subsection*{Example 2: $G=\alpha n^2/2$}

$G=\alpha n^2/2$ generate $\varphi \rightarrow \varphi +\alpha n$, which leads
$\partial_n \rightarrow \partial_n-\alpha \partial_\varphi$.  
$S=-G$ contributes to the canonicity condition (\ref{eq:canonicity 3}) and 
it is transformed as below.
\begin{align}
 &\langle \phi (q,p,\varphi,n ) | \frac{1}{i}\partial_n | \phi
 (q,p,\varphi,n ) \rangle =0  \notag \\
\rightarrow  &\langle \phi (q,p,\varphi,n ) |
 \frac{1}{i}\partial_{n^\prime} | \phi  (q,p,\varphi,n ) \rangle 
= \langle \phi (q,p,\varphi,n ) | \frac{1}{i}\partial_{n} +\alpha i
 \partial_\varphi | \phi  (q,p,\varphi,n ) \rangle \notag \\
=&\alpha n 
=- \frac{\partial S}{\partial n^\prime}. \label{eq:canonicity 3 w/ gauge-transformed frame}
\end{align}
One can see that the canonicity condition is satisfied after the gauge transformation.
The gauge transformation can be also discussed by transforming the
arguments of the state vector as below.
(Recall the discussion on the relation between the transformations of the c-numbers and those of the corresponding operators
in the last subsection.)
\begin{align}
 &\langle \phi (q,p,\varphi,n ) | \frac{1}{i}\partial_n | \phi
 (q,p,\varphi,n ) \rangle =0  \notag \\
\rightarrow  &\langle \phi (q,p,\varphi -\alpha n ,n ) |
 \frac{1}{i}\partial_{n} | \phi  (q,p,\varphi-\alpha n,n ) \rangle \notag \\
= & ( \langle \phi(q,p,\varphi,n )|  - \langle  \phi(q,p,\varphi,n ) |
 \cev{\partial}_\varphi\alpha n)
 \frac{1}{i}\partial_{n}  ( | \phi(q,p,\varphi,n ) \rangle   
- \alpha n \vec{\partial}_\varphi | \phi (q,p,\varphi,n )
 \rangle )\notag \\
=&\langle \phi (q,p,\varphi,n )| \frac{1}{i}\partial_{n}  | \phi (q,p,\varphi,n )\rangle 
-\alpha n \partial_\varphi \langle  \phi(q,p,\varphi,n ) |
 \frac{1}{i}\partial_{n}  | \phi (q,p,\varphi,n )\rangle \notag \\
&\hspace{15em}+ \alpha\langle \phi (q,p,\varphi,n )|  i\vec{\partial}_\varphi | \phi (q,p,\varphi,n )\rangle \notag \\
=&  \alpha n =- \frac{\partial S}{\partial n^\prime}. \label{eq:canonicity 3 w/ gauge-transformed state}
\end{align}
We have used 
$\langle \phi|\partial_\varphi\partial_n|\phi\rangle=\langle \phi|\partial_n\partial_\varphi|\phi\rangle$
and omitted higher-order infinitesimals.

\subsection*{Example 1: $G=\alpha pn$}
$G=\alpha pn$ generates the transformation
$(q,\varphi)\rightarrow (q^\prime,\varphi^\prime)=(q+\alpha n,\varphi+\alpha p)$.
In correspondence, the differential operators are transformed as 
$(\partial_p,\partial_n)\rightarrow(\partial_{p^\prime},\partial_{n^\prime})
=(\partial_{p}-\alpha \partial_\varphi,\partial_{n}-\alpha \partial_q)$.
Then, the canonicity conditions (\ref{eq:canonicity 3}) is transformed as
\begin{align}
 &\langle \phi (q,p,\varphi,n ) | \frac{1}{i}\partial_n | \phi
 (q,p,\varphi,n ) \rangle =0 
\notag \\
\rightarrow  &\langle \phi (q,p,\varphi,n ) |
 \frac{1}{i}\partial_{n^\prime} | \phi  (q,p,\varphi,n ) \rangle 
= \langle \phi (q,p,\varphi,n ) | \frac{1}{i}\partial_{n} +\alpha i
 \partial_q | \phi  (q,p,\varphi,n ) \rangle \notag \\
=&\alpha p 
=- \frac{\partial S}{\partial n^\prime}.
\end{align}
The canonicity condition (\ref{eq:canonicity 1}) is transformed as 
\begin{align}
 &\langle \phi (q,p,\varphi,n ) | \frac{1}{i}\partial_p | \phi
 (q,p,\varphi,n ) \rangle =0 
\notag \\
\rightarrow  &\langle \phi (q,p,\varphi,n ) |
 \frac{1}{i}\partial_{p^\prime} | \phi  (q,p,\varphi,n ) \rangle 
= \langle \phi (q,p,\varphi,n ) | \frac{1}{i}\partial_{p} +\alpha i
 \partial_\varphi | \phi  (q,p,\varphi,n ) \rangle \notag \\
 =&\alpha n 
=- \frac{\partial S}{\partial
 p^\prime}.\label{eq: canicity 1 transformed Ex 1}
\end{align}
$G=\alpha pn$ gives contributions of $O(p)$ and of 
$O(n)$
to the canonicity conditions (\ref{eq:canonicity 3}) and
(\ref{eq:canonicity 1}), respectively.

\subsection{Gauge symmetry breaking}
The transformation of the canonicity condition (\ref{eq:canonicity 3}) in Example 1 we have seen above,
\begin{align}
 \langle \phi (q,p,\varphi,n )| \frac{1}{i}\partial_n | \phi
 (q,p,\varphi,n ) \rangle =0 
\rightarrow  \langle \phi (q,p,\varphi,n ) |
 \frac{1}{i}\partial_{n^\prime} | \phi  (q,p,\varphi,n ) \rangle  =\alpha p,
\end{align}
implies that 
the canonical-variable condition (\ref{eq:can.var.cond 3}) 
\begin{align}
 \frac{i}{2}\langle \phi(q) | [\hat Q, \hat \Theta] | \phi (q) \rangle =　0,\tag{\ref{eq:can.var.cond 3}}
\end{align}
which is derived from the $O(p)$ term in the canonicity condition (\ref{eq:canonicity 3}),
should be changed by $\alpha$ by the gauge transformation.
However, this condition (\ref{eq:can.var.cond 3}) remains 0 after the gauge transformation as we
have seen in the beginning of Sect. 2.1.
This implies that the gauge symmetry in the canonicity conditions is 
broken by the adiabatic approximation. 
As we shall see below,
this is because we truncate the adiabatic expansion of $\hat G(q,p,n)$
to the first order and adopt $\hat G$ of the form $\hat G=p\hat Q +n\hat \Theta$.

The canonical-variable condition (\ref{eq:can.var.cond 3}) is also obtained
from the $O(n)$ term in the canonicity condition (\ref{eq:canonicity 1}).
As shown in Eq. (\ref{eq: canicity 1 transformed Ex 1}), 
the canonicity condition (\ref{eq:canonicity 1}) is changed by $\alpha
n$ by the gauge transformation 
and it is consistent with the discussion above.

Next we shall see the transformation of the canonicity condition (\ref{eq:canonicity 3})
in Example 2.
\begin{align}
 &\langle \phi (q,p,\varphi,n ) | \frac{1}{i}\partial_n | \phi
 (q,p,\varphi,n ) \rangle=0  
\notag \\
\rightarrow  &\langle \phi (q,p,\varphi,n ) |
 \frac{1}{i}\partial_{n^\prime} | \phi  (q,p,\varphi,n ) \rangle 
= \langle \phi (q,p,\varphi,n ) | \frac{1}{i}\partial_{n} +\alpha i
 \partial_\varphi | \phi  (q,p,\varphi,n ) \rangle \notag \\
& =\alpha n 
=- \frac{\partial S}{\partial n^\prime}.
\end{align}
This implies that the canonical-variable condition of $O(n)$ is changed
by $+\alpha$. However, unless we take into account higher-order terms in the 
expansion of $\hat G$,
there appears no term of $O(n)$ in the canonicity
condition (\ref{eq:canonicity 3}) and so it does not give a
canonical-variable condition of $O(n)$ [see (\ref{eq: ring
 Theta' 1st-order }) and (\ref{eq:2nd-order mathring Theta})].

\section{Expansion of $\hat G$ up to the second order}

Above we have seen that the gauge symmetry of the first-order canonical-variable
conditions is broken under the gauge transformations which are not point
transformations.
In this section, we show that the gauge symmetry is conserved if the
second-order terms
with respect to $p$ and $n$ in the generator $\hat G$ are taken into account.
We first present the basic definitions and equations in the case of
the second-order expansion of $\hat G$, and then discuss the gauge symmetry.
For details of the derivation of the basic equations, see Appendix A.

\subsection{Basic definitions and equations}
So far, we have taken into account
only the first-order terms in the expansion of $\hat G$.
Here we  consider $\hat G$ including the second-order terms as below.
\begin{equation}
\hat G(q,p,n) =p \hat Q^{(1)}(q)
+n \hat \Theta^{(1)}(q)+\frac{1}{2}p^2 \hat Q^{(2)}(q) +\frac{1}{2}n^2 \hat \Theta^{(2)} (q) +pn \hat X
\end{equation}
with
\begin{align}
\hat Q^{(i)}(q)&
=\sum_{\alpha > \beta}Q^{(i)}_{\alpha\beta}a_{\alpha}^\dagger a_{\beta}^\dagger  
+Q^{(i)*}_{\alpha\beta}a_{\beta}a_{\alpha}, \,\,\,\,(i=1,2),\notag \\
\hat \Theta^{(i)}(q)&
=\sum_{\alpha > \beta}\Theta^{(i)}_{\alpha\beta}a_{\alpha}^\dagger a_{\beta}^\dagger               
+\Theta^{(i)*}_{\alpha\beta}a_{\beta}a_{\alpha},  \,\,\,\,(i=1,2),\notag \\
\hat X(q)&=\sum_{\alpha > \beta}X_{\alpha\beta}a_{\alpha}^\dagger a_{\beta}^\dagger
                                   +X^{*}_{\alpha\beta}a_{\beta}a_{\alpha} .
\end{align}
These operators obey the time-reversal symmetry as follows. 
\begin{align}
\mathcal{T} \hat Q^{(1)}(q)\mathcal{T}^{-1} &=  \hat Q^{(1)}(q), \notag \\
\mathcal{T} \hat Q^{(2)}(q)\mathcal{T}^{-1} &= -\hat Q^{(2)}(q), \notag  \\
\mathcal{T} \hat \Theta^{(i)}(q)\mathcal{T}^{-1} &= -\hat
 \Theta^{(i)}(q), \,\,\,(i=1,2), \notag\\
\mathcal{T} \hat X(q)\mathcal{T}^{-1} &=  \hat X(q). 
\end{align}
The first-order operators in the previous section are denoted by $\hat Q=\hat
Q^{(1)}$ and $\hat \Theta=\hat \Theta^{(1)}$.
If we expand the collective Hamiltonian up to the second order, we obtain  
\begin{align}
\mathcal{H}(q,p,N) &=  V(q)+\frac{1}{2}B(q)p^2 +\lambda n
 +\frac{1}{2}D(q)n^2 \label{eq:2nd-order coll. H}
\end{align}
with
\begin{align}
 V(q)&=\langle\phi (q)|\hat H |\phi(q)\rangle ,\\
 B(q)&=\langle\phi (q)|[\hat H, i\hat  Q^{(2)}] |\phi(q)\rangle 
      -\langle\phi (q)|[[\hat H, \hat  Q^{(1)}],\hat Q^{(1)}] |\phi(q)\rangle, \\
 \lambda(q)&=\langle\phi (q)|[\hat H,i\hat \Theta^{(1)}] |\phi(q)\rangle, \\
 D(q)&=\langle\phi (q)|[\hat H, i\hat  \Theta^{(2)}] |\phi(q)\rangle 
      -\langle\phi (q)|[[\hat H, \hat  \Theta^{(1)}],\hat \Theta^{(1)}] |\phi(q)\rangle .
\end{align}
As discussed in Ref. \cite{Sato2015}, 
there appears no gauge symmetry if we employ
the collective Hamiltonian (\ref{eq:2nd-order coll. H}).
Therefore, we adopt the collective Hamiltonian up to $O(n)$:
\begin{align}
\mathcal{H}(q,p,N) &=  V(q)+\frac{1}{2}B(q)p^2 +\lambda n.  
\end{align}
The canonical-variable conditions are given as follows.

\noindent
\underline{ Zeroth-order canonical-variable conditions}
\begin{align}
&\langle \phi(q)|\,\hat P \,|\phi(q)\rangle =0.  \label{eq: 0th CVC with
 X 1}\\
&\langle \phi(q)|\hat Q^{(1)}|\phi(q)\rangle =0. \label{eq: 0th CVC with
 X 2}\\
&\langle \phi(q)|\,\hat \Theta^{(1)} \,|\phi(q)\rangle =0. \label{eq: 0th CVC
 with X 3}\\
&\langle \phi(q)|\,\tilde N \,|\phi(q)\rangle =0. \label{eq: 0th CVC
 with X 4}\\
&\langle \phi(q)|\,[\tilde N, \hat P] \,|\phi(q)\rangle =0. \label{eq:
 0th CVC with X 5}
\end{align}

\underline{ First-order canonical-variable conditions}
\begin{align}
&\langle \phi(q)|\,[\hat Q^{(1)}, \hat P] \,|\phi(q)\rangle =i. \label{eq:1st cvc with X 1}\\
&\langle \phi(q)|\,[\hat \Theta^{(1)}, \hat N] \,|\phi(q)\rangle =i.\label{eq:1st cvc with X 2} \\
&\langle \phi(q)|\,[\hat Q^{(1)}, \hat N] \,|\phi(q)\rangle =0.\label{eq:1st cvc with X 3} \\
&\langle \phi(q)|\,[\hat \Theta^{(1)},\hat P] \,|\phi(q)\rangle =0. \label{eq:1st cvc with X 4}\\
&\langle \phi(q)|\,\hat X+ \frac{i}{2}[\hat Q^{(1)}, \hat \Theta^{(1)}]
 \,|\phi(q)\rangle =0. \label{eq:1st cvc with X 5}\\
&\langle \phi(q)|\,\hat X+ \frac{i}{2}[\hat \Theta^{(1)}, \hat Q^{(1)}]
 \,|\phi(q)\rangle =0. \label{eq:1st cvc with X 6}\\
&\langle \phi(q)|\hat     Q^{(2)}|\phi(q)\rangle =0. \label{eq:1st cvc with X 7}\\
&\langle \phi(q)|\hat     \Theta^{(2)}|\phi(q)\rangle =0.\label{eq:1st cvc with X 8}
\end{align}
The conditions (\ref{eq:1st cvc with X 5}) and (\ref{eq:1st cvc with X
6})
can be rewritten as 
\begin{align}
&\langle \phi(q)|\,[\hat Q^{(1)}, \hat \Theta^{(1)}] \,|\phi(q)\rangle
 =0, 
\label{eq:1st cvc with X 5'}  
\\
&\langle \phi(q)|\,\hat X \,|\phi(q)\rangle =0. 
\label{eq:1st cvc with X 6'}
\end{align}
The moving-frame HFB \& QRPA equations are as below.

\noindent
\underline{Moving-frame HFB equation}
\begin{equation}
 \delta \langle \phi(q)|\hat H -\lambda \tilde N -\partial_q V\hat Q^{(1)} |\phi(q)\rangle =0.
\end{equation}

\noindent
\underline{ Moving-frame QRPA equations}
\begin{equation}
 \delta \langle \phi(q)|[\hat H-\lambda \tilde N , \hat Q^{(1)}] 
-\frac{1}{i}B(q)\hat P -\frac{1}{i}\partial_q V \hat Q^{(2)}
|\phi(q)\rangle =0. \label{eq:moving-frame QRPA1 with Q2}
\end{equation}

\begin{align}
 \delta \langle \phi(q)|
[\hat H -\lambda \tilde N -\partial_q V \hat Q^{(1)}, \frac{1}{i}\hat P]
-C(q)\hat Q^{(1)}-\partial_q \lambda \tilde N \hspace{16em}\notag\\
-\frac{1}{2B}
\partial_q V
\left\{[[\hat H-\lambda \tilde N,  \hat Q^{(1)}],\hat Q^{(1)}] -i [\hat H-\lambda \tilde N, \hat
 Q^{(2)}] 
-\frac{i}{2}\partial_q V [\hat Q^{(1)}, \hat Q^{(2)}]\right\}
|\phi(q)\rangle =0. \label{eq:moving-frame QRPA2 with Q2}
\end{align}
The moving-frame HFB equation remains unchanged when we include the second-order
operators $\hat Q^{(2)}$, $\hat \Theta^{(2)}$ and $\hat X$. In the moving-frame
QRPA equations of $O(p)$ and $O(p^2)$, $\hat Q^{(2)}$ is involved.

\subsection{Gauge symmetry in the case of the second-order expansion}
We shall investigate the gauge symmetry in the above case where we take into
account the second-order operators.
Although we are most interested in the non-point transformations,
we discuss not only Examples 1 and 2 but also Examples 3 and 4, for completeness.

\subsubsection{Gauge symmetry in Example 1}
We shall see how the operators are transformed under the
gauge transformation in Example 1.
It can be seen by transforming the arguments of the state vector
as $q\rightarrow q-\alpha n$ and $\varphi \rightarrow \varphi-\alpha p$.

\begin{align}
& e^{-i \varphi\tilde N} e^{i\hat G(q,p,n)}|\phi(q)\rangle \notag \\ 
\rightarrow &
e^{-i (\varphi-\alpha p)\tilde N} e^{i\hat G(q-\alpha n,
 p,n)}|\phi(q-\alpha n)\rangle \notag \\ 
=& e^{-i \varphi\tilde N} e^{i\alpha p\tilde N} e^{i\alpha n \hat P}
e^{i\hat G(q, p,n)}|\phi(q)\rangle \notag \\ 
=& e^{-i \varphi\tilde N} 
\exp\left\{
i\alpha p\tilde N+i\alpha n \hat P
+i \hat G
+\frac{1}{2}[i\alpha p\tilde N+i\alpha n \hat P, i \hat G]+\cdots
\right\}
|\phi(q)\rangle \notag \\ 
=& e^{-i \varphi\tilde N} 
\exp\left\{
ip(\hat Q^{(1)}+\alpha \tilde N )
+in(\hat \Theta^{(1)}+\alpha \hat P ) 
+i\frac{p^2}{2}(\hat Q^{(2)} +i\alpha [\tilde N, \hat Q^{(1)}] )
\right. \notag \\
&\hspace{0em} 
+i\frac{n^2}{2}(\hat \Theta^{(2)} +i \alpha [\hat P, \hat \Theta^{(1)}]
 )
\left.+ipn(\hat X 
+ \frac{i\alpha}{2}[\tilde N,\Theta^{(1)}]
+ \frac{i\alpha}{2}[\hat P,Q^{(1)}]
)
\right\}
|\phi(q)\rangle \label{eq:ex1 transformed state vec.}
\end{align}
Here we have used the Baker-Campbell-Hausdorff formula~\cite{Hall2003}:
\begin{align}
e^Xe^Y=\exp \left\{X+Y+\frac{1}{2}[X,Y] +\frac{1}{12}([X,[X,Y]]+[Y,[Y, X]]+\cdots \right\},
\end{align}
and omitted the second-order infinitesimals.
From Eq. (\ref{eq:ex1 transformed state vec.}), one can read the
following
transformation:
\begin{align}
\hat Q^{(1)} &\rightarrow  \hat Q^{(1)}      +\alpha \tilde N, \\ 
\hat \Theta^{(1)} &\rightarrow  \hat \Theta^{(1)}      +\alpha \hat P, \\ 
\hat Q^{(2)} &\rightarrow  \hat Q^{(2)}      + i \alpha [\tilde N, \hat Q^{(1)}], \\ 
\hat \Theta^{(2)} &\rightarrow  \hat \Theta^{(2)} + i \alpha [\hat P, \hat \Theta^{(1)}], \\ 
\hat X &\rightarrow \hat X +\frac{i\alpha }{2}([\tilde
 N,\Theta^{(1)}]+[\hat P, \hat Q^{(1)}]).
\end{align}

Let us see the gauge transformations of the canonical-variable
conditions involving the second-order operators.
The canonical-variable condition (\ref{eq:1st cvc with X 7}) which is derived
from the $O(p)$ term of 
the canonicity condition (\ref{eq:canonicity 1})
is transformed as
\begin{align}
 \langle \phi |\hat Q^{(2)} |\phi \rangle =0
\rightarrow
 \langle \phi |\hat Q^{(2)}    + i \alpha [\tilde N, \hat Q^{(1)}] |\phi \rangle =0,
\end{align}
so it is gauge invariant.
The canonical-variable condition (\ref{eq:1st cvc with X 8}) derived
from the $O(n)$ term of 
the canonicity condition (\ref{eq:canonicity 3})
is transformed as
\begin{align}
 \langle \phi |\hat \Theta^{(2)} |\phi \rangle =0
\rightarrow
 \langle \phi |\hat \Theta^{(2)}    + i \alpha [\hat P, \hat \Theta^{(1)}] |\phi \rangle =0,
\end{align}
so it is also gauge invariant. 
The canonical-variable condition (\ref{eq:1st cvc with X 5}) derived
from the $O(n)$ term of 
the canonicity condition (\ref{eq:canonicity 1})
is transformed as
\begin{align}
&\langle \phi(q)|\hat X + \frac{i}{2}[\hat Q^{(1)},\Theta^{(1)}]|\phi(q)\rangle =0\notag \\
\rightarrow & 
\langle \phi(q)|\hat X 
+\frac{i\alpha }{2}([\tilde
 N,\Theta^{(1)}]+[\hat P, \hat Q^{(1)}])
+ \frac{i}{2}[\hat Q^{(1)}+\alpha \tilde N ,\Theta^{(1)}+\alpha \hat P]|\phi(q)\rangle 
=\alpha.
\end{align}
It is changed by $\alpha$ as discussed in the previous section,
and thus the gauge symmetry is conserved.
It is easily seen  that the canonical-variable condition (\ref{eq:1st
cvc with X 6}), which is derived from the $O(p)$ term of 
the canonicity condition (\ref{eq:canonicity 3}), 
is also changed by $\alpha$ and that the gauge symmetry is not broken.
Note that the new operators after the gauge transformation
$\hat Q^{(1)\prime}=\hat Q^{(1)}+\alpha \tilde N$ and 
$\hat \Theta^{(1)\prime}=\hat \Theta^{(1)}+\alpha \hat P$ 
satisfy the weak canonical commutation relation,
$ \langle \phi|[\hat Q^{(1)\prime}, \hat \Theta^{(1)\prime}]|\phi\rangle=0.$

Next, we shall investigate the gauge symmetry of the moving-frame HFB \& QRPA
equations.
In the case of the expansion of $\hat G$ up to the first order,
if $[\tilde N,\hat Q^{(1)}]=0$,
the moving-frame HFB \& QRPA equations are invariant with the
transformation of the Lagrange multiplier:
\begin{align}
\lambda (q) & \rightarrow \lambda (q) -\alpha \partial_q V(q), \\
\partial_q \lambda (q) & \rightarrow \partial_q \lambda (q) -\alpha C(q).
\end{align}
However, the gauge symmetry of the moving-frame QRPA equations is
actually broken because, as easlily confirmed, 
the commutator $[\tilde N,\hat Q^{(1)}]$ is non-zero.

Because $\hat Q^{(2)}$ appears only in the moving-frame QRPA equations,
the moving-frame HFB equation remains gauge invariant if we take into
account the second-order operators.
The moving-frame QRPA equation of $O(p)$ (\ref{eq:moving-frame QRPA1 with Q2}) is transformed as
\begin{align}
&\hspace{0em} 
\delta \langle \phi(q)|
[\hat H-\lambda \tilde N -\partial_q V \hat Q^{(1)}, \hat Q^{(1)}] 
-\frac{1}{i}B(q)\hat P -\frac{1}{i}\partial_q V 
\hat Q^{(2)} |\phi(q)\rangle =0,\notag\\
 \rightarrow &
\delta \langle \phi(q)|
[\hat H-\lambda \tilde N -\partial_q V \hat Q^{(1)}, \hat Q^{(1)}+\alpha
 \tilde N ] 
-\frac{1}{i}B(q)\hat P \notag \\
&\hspace{12em} -\frac{1}{i}\partial_q V 
(\hat Q^{(2)} + i \alpha [\tilde N, \hat Q^{(1)}])|\phi(q)\rangle =0,\notag \\
\Leftrightarrow & \delta \langle \phi(q)|
[\hat H-\lambda \tilde N -\partial_q V \hat Q^{(1)}, \hat Q^{(1)}] 
-\frac{1}{i}B(q)\hat P -\frac{1}{i}\partial_q V 
\hat Q^{(2)} |\phi(q)\rangle =0,
\end{align}
and thus it is gauge invariant.
Without $Q^{(2)}$, the moving-frame QRPA equation of $O(p)$ is not
gauge invariant because $[\tilde N, \hat Q^{(1)}] \neq 0$ .
With $Q^{(2)}$ included, it is gauge invariant even if $[\tilde N, \hat
Q^{(1)}]\neq 0$.
Thus one can see that the gauge symmetry breaking in
the moving-frame QRPA equation of $O(p)$
is because of the truncation of the adiabatic expansion of $\hat G$
to the first order.

As shown below, 
the moving-frame QRPA equation of $O(p^2)$ 
(\ref{eq:moving-frame QRPA2 with Q2})
is not
gauge invariant.
As in the case of the first-order expansion, the first three terms in
Eq. (\ref{eq:moving-frame QRPA2 with Q2})
\begin{equation}
[\hat H -\lambda \tilde N -\partial_q V \hat Q^{(1)}, \frac{1}{i}\hat P]
-C(q)\hat Q^{(1)}-\partial_q \lambda \tilde N 
\end{equation}
are gauge invariant. 
In the case of the first-order expansion, the fourth term 
\begin{equation}
 -\frac{1}{2B}
\partial_q V
\left\{[[\hat H-\lambda \tilde N,  \hat Q^{(1)}],\hat Q^{(1)}]\right\}
\end{equation}
is not gauge invariant, and $[\tilde N, \hat Q^{(1)}]=0$ is required for
the gauge symmetry.
Therefore,
we only have to check the gauge invariance of the part
\begin{equation}
[[\hat H-\lambda \tilde N,  \hat Q^{(1)}],\hat Q^{(1)}] -i [\hat H-\lambda \tilde N, \hat
 Q^{(2)}] 
-\frac{i}{2}\partial_q V [\hat Q^{(1)}, \hat Q^{(2)}].
\end{equation}
After a lengthy calculation,
one finds that it transforms as
\begin{align}
 &[[\hat H_M,  \hat Q^{(1)}],\hat Q^{(1)}] -i[\hat H -\lambda \tilde N , \hat Q^{(2)}] -\frac{i}{2}\partial_q V[\hat Q^{(1)}, \hat Q^{(2)}] \notag \\
\rightarrow
 &[[\hat H_M,  \hat Q^{(1)}],\hat Q^{(1)}] -i[\hat H -\lambda \tilde N , \hat Q^{(2)}] -\frac{i}{2}\partial_q V[\hat Q^{(1)}, \hat Q^{(2)}] \notag \\
&+\frac{1}{2}\alpha \partial_q V[ [\tilde N, \hat Q^{(1)}],\hat Q^{(1)}]
-\frac{3}{2}i\alpha \partial_q V[\tilde N, \hat Q^{(2)}]
 +\frac{1}{2}\alpha^2\partial_q V[\tilde N, [\tilde N, \hat Q^{(1)}]].
\label{eq:gauge-dep part in mfQRPA2 with Q2}
\end{align}
Thus the moving-frame QRPA equation of $O(p^2)$ is not gauge invariant
because $[ [\tilde N, \hat Q^{(1)}],\hat Q^{(1)}]$ and $[\tilde N, \hat Q^{(2)}]$
do not vanish.
In fact, it is gauge invariant if we take into account the third-order
operator $Q^{(3)}$, as we shall see in the next section.
To sum up, with $\hat Q^{(2)}$ included, 
all the ASCC equations and conditions of $O(1)$ and $O(p)$ are invariant
under the gauge transformation in Example 1.
The moving-frame QRPA equation of $O(p^2)$ is not gauge invariant. 

\subsubsection{Gauge symmetry in Example 2}
In Example 2, the generator $G=\frac{\alpha}{2}n^2$ generates $\varphi\rightarrow \varphi
+\alpha n$.
By transforming the argument of the state vector as $\varphi \rightarrow
\varphi -\alpha n$, 
\begin{align}
& e^{-i\varphi \tilde N}e^{i\hat G}|\phi(q)\rangle \notag \\
\rightarrow &  e^{-i(\varphi -\alpha n) \tilde N}e^{iG}|\phi(q)\rangle
=   e^{-i\varphi \tilde N}e^{i\alpha n \tilde N}e^{iG}|\phi(q)\rangle \notag \\
= &  e^{-i\varphi \tilde N}\exp\left\{
i p\hat Q^{(1)} + in \Theta^{(1)}+\frac{i}{2}p^2\hat Q^{(2)}+\frac{i}{2}n^2\hat
 \Theta^{(2)}+ pn\hat X+i\alpha n \tilde N \right. \notag \\
&\left. \hspace{6em}+\frac{1}{2}[i\alpha n \tilde N, i p\hat Q^{(1)} + in \Theta^{(1)}] +\cdots
 \right\}|\phi(q)\rangle \notag \\
= &  e^{-i\varphi \tilde N}\exp\left\{
i p\hat Q^{(1)} + in (\Theta^{(1)}+\alpha \tilde N) +\frac{i}{2}p^2\hat
 Q^{(2)} \right.\notag \\
&\left. +\frac{i}{2}n^2( \hat  \Theta^{(2)}+i\alpha[\tilde N, \hat\Theta^{(1)}] )
+ pn(\hat X+\frac{i}{2}\alpha  [\tilde N,  \hat Q^{(1)} ]) +\cdots
 \right\}|\phi(q)\rangle,
\end{align}
we obtain 
\begin{align}
&\hat \Theta^{(1)}\rightarrow \hat \Theta^{(1)}+ \alpha \tilde N, \\
&\hat \Theta^{(2)}\rightarrow \hat \Theta^{(2)}+ i\alpha [\tilde N, \hat
 \Theta^{(1)}], \\
&\hat X \rightarrow \hat X + \frac{i}{2}\alpha  [\tilde N, \hat Q^{(1)}] .
\end{align}
Because $\hat Q^{(2)}$ is not transformed, the moving-frame HFB \& QRPA
equations are invariant as in the case of the first-order expansion of
$\hat G$. 
The canonical-variable conditions involving the second-order 
operators are transformed as
\begin{align}
&\langle \phi (q) |\hat \Theta^{(2)} |\phi(q)\rangle \rightarrow \langle \phi (q) |\hat
 \Theta^{(2)} |\phi(q)\rangle + i\alpha \langle \phi (q) |[\tilde N,
 \hat \Theta^{(1)}] |\phi(q)\rangle=\alpha \label{eq: can.var. Theta(2) eg2}　\\
&\langle \phi (q) |\hat X |\phi(q)\rangle \rightarrow \langle \phi (q)
 |\hat X |\phi(q)\rangle+ \frac{i}{2}\alpha \langle \phi (q) | [\tilde
 N, \hat Q^{(1)}] |\phi(q)\rangle=0.
\label{eq: can.var. X eg2}
\end{align}
Here we employ the conditions (\ref{eq:1st cvc with X 5'}) and
(\ref{eq:1st cvc with X 6'})
instead of (\ref{eq:1st cvc with X 5}) and (\ref{eq:1st cvc with X
6}).
The other canonical-variable conditions are invariant under the gauge transformation.
Among the canonical-variable conditions, most noteworthy is (\ref{eq: can.var. Theta(2) eg2}). 
$\hat G=\alpha n^2/2$ gives a contribution of $\alpha n$ 
to the gauge transformation of the canonicity condition (\ref{eq:canonicity 3}).
The transformation (\ref{eq: can.var. Theta(2) eg2}) correctly 
reflects the gauge transformation of the canonicity condition (\ref{eq:canonicity 3}).

\subsubsection{Gauge symmetry in Example 3}

$G=\alpha qn$ generates 
$(\varphi,p) \rightarrow (\varphi+\alpha q,p - \alpha n)$.
By considering the transformation,
\begin{align}
& e^{-i\varphi \tilde N}e^{i\hat G}|\phi(q)\rangle \notag \\
\rightarrow &  e^{-i(\varphi -\alpha q) \tilde N}
\exp \left\{ i\left[ (p+\alpha n)\hat Q^{(1)} 
+ \frac{1}{2}(p+\alpha n)^2 \hat Q^{(2)} \right.\right.\notag\\
&\hspace{4em}\left. \left. +n \hat \Theta^{(1)} + \frac{1}{2}n^2 \hat \Theta^{(2)}+ (p+\alpha
 n)n\hat X	      \right]\right\}|\phi(q)\rangle  \notag \\
& = e^{-i\varphi\tilde N}e^{i\alpha q \tilde N }e^{-i q\hat P}
\exp \left\{ i\left[ p\hat Q^{(1)} +n (\hat \Theta^{(1)}+\alpha \hat Q^{(1)}) 
+ \frac{1}{2}p^2 \hat Q^{(2)} \right.\right.\notag\\
&\hspace{4em}\left. \left.  + \frac{1}{2}n^2 (\hat \Theta^{(2)}+2\alpha
 \hat X+\alpha^2 \hat Q^{(2)})+ pn(\hat X+\alpha \hat Q^{(2)})	      \right]\right\}|_{q=0}|\phi(0)\rangle  \notag \\
& = e^{-i\varphi\tilde N}e^{-i q(\hat P-\alpha \tilde N)}
\exp \left\{ i\left[ p\hat Q^{(1)} +n (\hat \Theta^{(1)}+\alpha \hat Q^{(1)}) 
+ \frac{1}{2}p^2 \hat Q^{(2)} \right.\right.\notag\\
&\hspace{4em}\left. \left.  + \frac{1}{2}n^2 (\hat \Theta^{(2)}+2\alpha
 \hat X)+ pn(\hat X+\alpha \hat Q^{(2)})	      \right]\right\}|_{q=0}|\phi(0)\rangle,  
\end{align}
we find
\begin{align}
& \hat P \rightarrow \hat P - \alpha \tilde N, \\
&\hat \Theta^{(1)}\rightarrow \hat \Theta^{(1)}+ \alpha\hat Q^{(1)}, \\
&\hat \Theta^{(2)}\rightarrow \hat \Theta^{(2)}+ 2\alpha \hat X, \\
&\hat X \rightarrow \hat X + \alpha\hat Q^{(2)}.
\end{align}
The inclusion of the second-order operators does not affect the discussion in Ref. \cite{Sato2015}.
While the moving-frame HFB equation is gauge invariant, the moving-frame
QRPA equation of $O(p)$ is not.
The moving-frame QRPA equation of $O(p^2)$ is not gauge invariant
because $[\tilde N, \hat Q^{(1)}]\neq 0$.
As easily seen, the canonical-variable conditions are all gauge invariant.

\subsubsection{Gauge symmetry in Example 4}

$G=\alpha \varphi n$ generates
$\varphi \rightarrow (1+\alpha)\varphi = e^\alpha \varphi$, 
$n \rightarrow (1-\alpha)n = e^{-\alpha} n$.
By considering the transformation,
\begin{align}
& e^{-i\varphi \tilde N}e^{i\hat G}|\phi (q) |\phi (q) \rangle \notag \\
\rightarrow &e^{-i(1-\alpha )\varphi\tilde N } 
\exp\left\{ 
i p\hat Q^{(1)} + i(1+\alpha)n \hat \Theta^{(1)}+\frac{i}{2}p^2\hat
 Q^{(2)}\right.\notag \\
&\hspace{4em}\left. +\frac{i}{2}(1+\alpha )^2n^2\hat
 \Theta^{(2)}+ p(1+\alpha)n\hat X
\right\}|\phi (q) \rangle \notag \\
= &e^{-i\varphi (1-\alpha )\tilde N } 
\exp\left\{ 
i p\hat Q^{(1)} + in(1+\alpha) \Theta^{(1)}+\frac{i}{2}p^2\hat
 Q^{(2)}\right.\notag \\
&\hspace{4em}\left. +\frac{i}{2}n^2(1+2\alpha )\hat
 \Theta^{(2)}+ pn(1+\alpha)\hat X
\right\}|\phi (q) \rangle, 
\end{align}
we find
\begin{align}
&\hat \Theta^{(1)}\rightarrow (1+\alpha )\hat \Theta^{(1)}= e^{\alpha }\Theta^{(1)}, \\
&\hat \Theta^{(2)}\rightarrow (1+2\alpha )\hat \Theta^{(2)} = e^{2\alpha }\Theta^{(2)}, \\
&\hat X \rightarrow (1+\alpha)\hat X=e^{\alpha}\hat X, \\
&\tilde N \rightarrow (1-\alpha )\tilde N= e^{-\alpha }\tilde N. 
\end{align}
Also in this case, the inclusion of the second-order operators does not affect the discussion 
in Ref. \cite{Sato2015}.
It is clear that the moving-frame HFB \& QRPA equations and the
canonical-variable conditions are gauge invariant.

\section{Expansion of $\hat G$ up to the third order}
In this section, we consider the expansion of $\hat G$ up to the third order.
(For the derivation of the basic equations in this section, see Appendix A.)
$\hat G$ is expanded as
\begin{align}
\hat G(q,p,n) &=p \hat Q^{(1)}(q)+n \hat \Theta^{(1)}(q)
+\frac{1}{2}p^2 \hat Q^{(2)}(q) 
+\frac{1}{2}n^2 \hat \Theta^{(2)} (q) +pn \hat X \notag\\
&+\frac{1}{3!}p^3 \hat Q^{(3)}(q) 
+\frac{1}{3!}n^3 \hat \Theta^{(3)}(q) 
+\frac{1}{2}p^2n \hat O^{(2,1)}(q) 
+\frac{1}{2}pn^2 \hat O^{(1,2)}(q) .
\end{align}
The time-reversal symmetry of these operators are as follows.
\begin{align}
\mathcal{T} \hat Q^{(i)}(q)\mathcal{T}^{-1} &=(-1)^{(i-1)}\hat Q^{(i)}(q),\,\,\,(i=1,2,3), \notag \\
\mathcal{T} \hat \Theta^{(i)}(q)\mathcal{T}^{-1} &= -\hat
 \Theta^{(i)}(q), \hspace{4em}(i=1,2,3),\notag\\
\mathcal{T} \hat X(q)\mathcal{T}^{-1} &=  \hat X(q), \notag\\
\mathcal{T} \hat O^{(2,1)}(q)\mathcal{T}^{-1} &=  -\hat O^{(2,1)}(q), \notag \\
\mathcal{T} \hat O^{(1,2)}(q)\mathcal{T}^{-1} &=  \hat O^{(1,2)}(q). \notag 
\end{align}
The third-order operators do not appear in the zeroth- and first-order
canonical-variable conditions.
They are involved in the second-order canonical-variable conditions:

\noindent
\underline{ Second-order canonical-variable conditions}
\begin{align}
&\frac{1}{2}\langle \phi(q)|\,\hat Q^{(3)}-\frac{i}{2}[\hat Q^{(1)},
 \hat     Q^{(2)}]\,|\phi(q)\rangle =0, \label{eq:2nd CVC w Q3 1}\\
&\frac{1}{2}\langle \phi(q)|\,\hat \Theta^{(3)}-\frac{i}{2}[ \hat    \Theta^{(1)}, \hat \Theta^{(2)}]\, |\phi(q)\rangle =0, \label{eq:2nd CVC w Q3 2}\\
&\frac{1}{2}\langle \phi(q)|\,\hat O^{(2,1)}
+\frac{i}{2} [\hat\Theta^{(1)}, \hat Q^{(2)}]
 -\frac{1}{3}[[\hat \Theta^{(1)}, \hat Q^{(1)}], \hat Q^{(1)}]
+i[\hat X,\hat Q^{(1)}]  
\,|\phi(q)\rangle =0 ,\label{eq:2nd CVC w Q3 3}\\
&\frac{1}{2}\langle \phi(q)|\,\hat O^{(1,2)}+
\frac{i}{2}[\hat Q^{(1)},\hat \Theta^{(2)}]
 -\frac{1}{3}[[\hat Q^{(1)}, \hat \Theta^{(1)}], \hat \Theta^{(1)}]
+i[\hat X,\hat \Theta^{(1)}] \,|\phi(q)\rangle =0 ,\label{eq:2nd CVC w Q3 4}\\
&\langle \phi(q)|\,
\hat O^{(2,1)}
+\frac{i}{2}[\hat Q^{(2)},\hat \Theta^{(1)}]  +\frac{1}{6}[[ \hat \Theta^{(1)},\hat Q^{(1)}] ,\hat Q^{(1)}]
|\phi(q)\rangle =0 ,\label{eq:2nd CVC w Q3 5}\\
&\langle \phi(q)|\,
\hat O^{(1,2)}
+\frac{i}{2}[\hat \Theta^{(2)}, \hat Q^{(1)}]
+\frac{1}{6}[[\hat Q^{(1)},\hat \Theta^{(1)}] ,\hat \Theta^{(1)}]
\,|\phi(q)\rangle =0 ,\label{eq:2nd CVC w Q3 6}\\
&\frac{1}{2}\langle \phi(q)|(\,i[ \hat P, \hat Q^{(2)}] -[[ \hat P, \hat Q^{(1)}], \hat Q^{(1)}]\,)|\phi(q)\rangle =0 ,\label{eq:2nd CVC w Q3 7}\\
&\frac{1}{2}\langle \phi(q)|(\,i[ \hat P, \hat \Theta^{(2)}] -[[ \hat P,
 \hat \Theta^{(1)}],\hat \Theta^{(1)}]\,)|\phi(q)\rangle =0,
 \label{eq:2nd CVC w Q3 8}\\
&\frac{1}{2}\langle \phi(q)|(\,i[ \hat N, \hat Q^{(2)}] -[[ \hat N, \hat Q^{(1)}], \hat Q^{(1)}]\,)|\phi(q)\rangle =0, \label{eq:2nd CVC w Q3 9}\\
&\frac{1}{2}\langle \phi(q)|(\,i[ \hat N, \hat \Theta^{(2)}] -[[ \hat N, \hat \Theta^{(1)}],\hat \Theta^{(1)}]\,)|\phi(q)\rangle =0, \label{eq:2nd CVC w Q3 10}\\ 
&\langle \phi(q)|\,i[\hat P,\hat X]-\frac{1}{2}(\,[[\hat P , \hat Q^{(1)}],\hat \Theta^{(1)}]+[[\hat P , \hat \Theta^{(1)}],\hat Q^{(1)}]\, )|\phi(q)\rangle =0, \label{eq:2nd CVC w Q3 11}\\
&\langle \phi(q)|\,i[\hat N,\hat X]-\frac{1}{2}([[\hat N , \hat
 Q^{(1)}],\hat \Theta^{(1)}]+[[\hat N , \hat \Theta^{(1)}],\hat
 Q^{(1)}]\, )|\phi(q)\rangle =0. \label{eq:2nd CVC w Q3 12}
\end{align}
The second-order canonical-variable conditions are gauge invariant if
the expansion of $\hat G$ up to the third order is taken into account.
The conditions (\ref{eq:2nd CVC w Q3 3})-(\ref{eq:2nd CVC w Q3 6}) can be rewritten as follows.
\begin{align}
&\langle \phi(q)|\,2\hat O^{(2,1)}
 -\frac{1}{6}[[\hat \Theta^{(1)}, \hat Q^{(1)}], \hat Q^{(1)}]
+i[\hat X,\hat Q^{(1)}]  
\,|\phi(q)\rangle =0, \\
&\frac{1}{2}\langle \phi(q)|\,
i[\hat \Theta^{(1)},\hat Q^{(2)}]
-\frac{1}{2}[[ \hat \Theta^{(1)},\hat Q^{(1)}] ,\hat Q^{(1)}]
+i[\hat X,\hat Q^{(1)}]
|\phi(q)\rangle =0 ,\\
&\langle \phi(q)|\,2\hat O^{(1,2)}
-\frac{1}{6}[[\hat Q^{(1)}, \hat \Theta^{(1)}], \hat \Theta^{(1)}]
+i[\hat X,\hat \Theta^{(1)}] 
\,|\phi(q)\rangle =0,\\
&\frac{1}{2}\langle \phi(q)|\,
i [\hat Q^{(1)}, \hat \Theta^{(2)}]
-\frac{1}{2}[[\hat Q^{(1)},\hat \Theta^{(1)}] ,\hat \Theta^{(1)}]
+i[\hat X,\hat \Theta^{(1)}] 
\,|\phi(q)\rangle =0.
\end{align}

We shall consider the gauge transformation of operators in Example 1.
By transforming the argument of the state vector as
$q \rightarrow q -\alpha n$  and $\varphi \rightarrow \varphi -\alpha
p$,
\begin{align}
 &e^{-i\varphi\tilde N}e^{i\hat G(q , p, n)}|\phi(q)\rangle \notag \\
\rightarrow
&e^{-i(\varphi-\alpha p)\tilde N}e^{i\hat G(q-\alpha n , p, n)}|\phi (q-\alpha n)\rangle 
=e^{-i\varphi\tilde N}e^{i\alpha p\tilde N +i\alpha n \hat P}e^{i\hat G(q , p, n)}|\phi (q)\rangle \notag \\
=&e^{-i\varphi\tilde N}\exp \left\{
i\hat G + i\alpha (p\tilde N +n\hat P) +\frac{1}{2}[ i\alpha (p\tilde N
 +n\hat P), i\hat G] 
+\frac{1}{12}[i\hat G, [i\hat G,  i\alpha (p\tilde N +n\hat P)]]
\right\} |\phi(q) \rangle  \notag \\
=&e^{-i\varphi\tilde N}
\exp \left\{
ip (\hat Q^{(1)}     +\alpha \tilde N)
+in (\hat \Theta^{(1)}+\alpha \hat P) \right.\notag \\
&+\frac{i}{2}p^2 (\hat Q^{(2)}      +i\alpha [\tilde N, \hat Q^{(1)}]) 
+\frac{i}{2}n^2 (\hat \Theta^{(2)} +i\alpha [\hat P  , \hat \Theta^{(1)}]) 
+ipn (\hat X +\frac{i}{2}[\tilde N, \hat \Theta^{(1)}] +\frac{i}{2}[\hat P, \hat Q^{(1)}])
\notag \\
&+\frac{i}{3!}p^3 (\hat Q^{(3)}     +\frac{3}{2}\alpha i [\tilde N, \hat Q^{(2)}]      -\frac{1}{2}\alpha [\hat Q^{(1)},     [\hat Q^{(1)}, \tilde N]]) \notag \\
&+\frac{i}{3!}n^3 (\hat \Theta^{(3)}+\frac{3}{2}\alpha i [\hat   P, \hat \Theta^{(2)}] -\frac{1}{2}\alpha [\hat \Theta^{(1)},[\hat \Theta^{(1)}, \hat P]]) \notag \\
&+\frac{i}{2}pn^2 \left(\hat O^{(1,2)} +\frac{i}{2}\alpha [\tilde N, \hat \Theta^{(2)}]+i\alpha [\hat P, \hat X] \right.\notag\\
&\left.-\frac{1}{6}\alpha ([\hat Q^{(1)}, [\hat \Theta^{(1)}, \hat P]]+[\hat \Theta^{(1)}, [\hat Q^{(1)}, \hat P]]+[\hat \Theta^{(1)},[\hat \Theta^{(1)},\tilde N]] ))
\right)\notag\\
&+\frac{i}{2}p^2n \left(\hat O^{(2,1)}(q) +\frac{i}{2}\alpha [\hat P, \hat Q^{(2)}]+i\alpha [\tilde N, \hat X] \right.\notag\\
&\left.\left.-\frac{1}{6}\alpha ([\hat \Theta^{(1)}, [\hat Q^{(1)}, \tilde N]]+[\hat Q^{(1)}, [\hat \Theta^{(1)}, \tilde N]]+[\hat Q^{(1)},[\hat Q^{(1)},\hat P]] ))
\right)
\right\} |\phi(q) \rangle, 
\end{align}
we find
\begin{align}
&\hat Q^{(1)} \rightarrow \hat Q^{(1)}     +\alpha \tilde N, \\
&\hat \Theta^{(1)} \rightarrow \hat \Theta^{(1)}+\alpha \hat P, \\
&\hat Q^{(2)} \rightarrow \hat Q^{(2)}      +i\alpha [\tilde N, \hat Q^{(1)}], \\
&\hat \Theta^{(2)} \rightarrow \hat \Theta^{(2)} +i\alpha [\hat P  , \hat \Theta^{(1)}] ,\\
&\hat X \rightarrow \hat X +\frac{i}{2}[\tilde N, \hat \Theta^{(1)}] +\frac{i}{2}[\hat P, \hat Q^{(1)}] ,\\
&\hat Q^{(3)} \rightarrow \hat Q^{(3)}     +\frac{3}{2}\alpha i [\tilde N, \hat Q^{(2)}]      -\frac{1}{2}\alpha [\hat Q^{(1)},     [\hat Q^{(1)}, \tilde N]], \\
&\hat \Theta^{(3)} \rightarrow \hat \Theta^{(3)}+\frac{3}{2}\alpha i [\hat   P, \hat \Theta^{(2)}] -\frac{1}{2}\alpha [\hat \Theta^{(1)},[\hat \Theta^{(1)}, \hat P]], \\
&\hat O^{(1,2)} \rightarrow \hat O^{(1,2)} +\frac{i}{2}\alpha [\tilde N, \hat \Theta^{(2)}]+i\alpha [\hat P, \hat X] \notag\\
&\hspace{4em}-\frac{1}{6}\alpha ([\hat Q^{(1)}, [\hat \Theta^{(1)}, \hat P]]+[\hat \Theta^{(1)}, [\hat Q^{(1)}, \hat P]]+[\hat \Theta^{(1)},[\hat \Theta^{(1)},\tilde N]] ), 
\end{align}
\begin{align}
&\hat O^{(2,1)} \rightarrow \hat O^{(2,1)}(q) +\frac{i}{2}\alpha [\hat P, \hat Q^{(2)}]+i\alpha [\tilde N, \hat X] \notag\\
&\hspace{4em}-\frac{1}{6}\alpha ([\hat \Theta^{(1)}, [\hat Q^{(1)}, \tilde N]]+[\hat Q^{(1)}, [\hat \Theta^{(1)}, \tilde N]]+[\hat Q^{(1)},[\hat Q^{(1)},\hat P]] ).
\end{align}
As one can ascertain easily,
the second-order canonical-variable conditions (\ref{eq:2nd CVC w Q3 1})-(\ref{eq:2nd CVC w Q3 12}) are invariant under this transformation.
For example, in the case of
the  $O(p^2)$ canonical-variable condition (\ref{eq:2nd CVC w Q3 1}),
\begin{align}
& \langle \phi(q)|  \hat Q^{(3)} -\frac{i}{2}[\hat Q^{(1)},\hat Q^{(2)}] |\phi(q)\rangle =0\notag \\
\rightarrow 
& \langle \phi(q)|  \hat Q^{(3)}  +\frac{3}{2}\alpha i [\tilde N, \hat Q^{(2)}]      -\frac{1}{2}\alpha [\hat Q^{(1)},     [\hat Q^{(1)}, \tilde N]] 
-\frac{i}{2}[\hat Q^{(1)}+\alpha \tilde N,\hat Q^{(2)} +i\alpha [\tilde N, \hat Q^{(1)}]] |\phi(q)\rangle \notag \\
=& \langle \phi(q)|  \hat Q^{(3)} -\frac{i}{2}[\hat Q^{(1)},\hat
 Q^{(2)}] |\phi(q)\rangle 
+\alpha \langle \phi(q)|  i[\tilde N , \hat Q^{(2)}] -  [\hat Q^{(1)}, [\hat Q^{(1)},\tilde N] ]|\phi(q)\rangle \notag \\
=&0.
\end{align}
It is clear that
this canonical-variable condition is not gauge invariant unless
$\hat Q^{(3)}$ is taken into account.

We shall move on to the gauge symmetry of the moving-frame QRPA equation of
$O(p^2)$.
The moving-frame HFB \& QRPA equations are given as follows.

\noindent
\underline{Moving-frame HFB equation}
\begin{equation}
 \delta \langle \phi(q)|\hat H -\lambda \tilde N -\partial_q V\hat Q^{(1)} |\phi(q)\rangle =0.
\end{equation}

\noindent
\underline{ Moving-frame QRPA equations}
\begin{equation}
 \delta \langle \phi(q)|[\hat H-\lambda \tilde N , \hat Q^{(1)}] 
-\frac{1}{i}B(q)\hat P -\frac{1}{i}\partial_q V \hat Q^{(2)} |\phi(q)\rangle =0.
\label{eq:moving-frame QRPA1 with Q3}
\end{equation}

\begin{align}
 \delta \langle \phi(q)|
[\hat H -\lambda \tilde N -\partial_q V \hat Q^{(1)}, \frac{1}{i}\hat P]
-C(q)\hat Q^{(1)}-\partial_q \lambda \tilde N \hspace{18em} \notag\\
-\frac{1}{2B}
\partial_q V
\left\{[[\hat H-\lambda \tilde N,  \hat Q^{(1)}],\hat Q^{(1)}] -i [\hat H-\lambda \tilde N, \hat
 Q^{(2)}] 
+\partial_q V( \hat Q^{(3)}-\frac{i}{2} [\hat Q^{(1)}, \hat Q^{(2)}])\right\}
|\phi(q)\rangle =0. \label{eq:moving-frame QRPA2 with Q3 }
\end{align}
As discussed in the previous section, the first three terms are
gauge invariant in Eq. (\ref{eq:moving-frame QRPA2 with Q3 }).
We only have to check the gauge symmetry of the rest.
With the  use of the previous result (\ref{eq:gauge-dep part in mfQRPA2
with Q2}), 
we easily obtain
\begin{align}
&[[\hat H-\lambda \tilde N,  \hat Q^{(1)}],\hat Q^{(1)}] -i [\hat H-\lambda \tilde N, \hat
 Q^{(2)}] + \partial_q V( \hat Q^{(3)}-\frac{i}{2} [\hat Q^{(1)}, \hat Q^{(2)}]) \notag \\
\rightarrow & [[\hat H-\lambda \tilde N,  \hat Q^{(1)}],\hat Q^{(1)}] -i [\hat H-\lambda \tilde N, \hat
 Q^{(2)}]  -\frac{i}{2} \partial_q V [\hat Q^{(1)}, \hat Q^{(2)}] \notag\\
&\hspace{4em}+\frac{1}{2}\alpha \partial_q V[ [\tilde N, \hat Q^{(1)}],\hat Q^{(1)}]
-\frac{3}{2}i\alpha \partial_q V[\tilde N, \hat Q^{(2)}] \notag \\
& \hspace{6em}+ \partial_q V( \hat Q^{(3)}     +\frac{3}{2}\alpha i [\tilde N, \hat Q^{(2)}]      -\frac{1}{2}\alpha [\hat Q^{(1)},     [\hat Q^{(1)}, \tilde N]] )\notag \\
=& [[\hat H-\lambda \tilde N,  \hat Q^{(1)}],\hat Q^{(1)}] -i [\hat H-\lambda \tilde N, \hat
 Q^{(2)}]  -\frac{i}{2} \partial_q V [\hat Q^{(1)}, \hat Q^{(2)}] 
 + \partial_q V\hat Q^{(3)}.
\end{align}
Thus, the moving-frame QRPA equation of $O(p^2)$ is gauge invariant.
As discussed in Ref. \cite{Sato2015},
it is no trivial that the moving-frame QRPA equation of $O(p^2)$ is
gauge invariant if the third-order operator is included.
At the level of the equation of collective submanifold, 
the gauge symmetry of the equation of motion is conserved.
Specifically, with the transformation of the Lagrange multiplier
\begin{equation}
\lambda \rightarrow \lambda -\alpha \partial_q V -\frac{1}{2}\alpha \partial_q B p^2
\label{eq:lambda' with p2 in Example 1}
\end{equation}
the equation of collective submanifold is gauge invariant.
The change of the Lagrange multiplier $\lambda$ under the gauge
transformation (\ref{eq:lambda' with p2 in Example 1}) 
contains a term of $O(p^2)$. 
Therefore,
when the equation of collective submanifold is divided into the three
equations depending on the order of $p$ as shown in Eqs. (\ref{eq:moving-frame HFB in App. B})-(\ref{eq:p^2 EoCSM}),
the equation of $O(p^2)$ is not gauge invariant.
The gauge invariance of the moving-frame QRPA equation of $O(p^2)$ can
be attributed by the fact that it is derived by using both of the 
$O(1)$ and $O(p^2)$ expansions of the collective submanifold.

\section{Concluding remarks}
In this paper, we have analyzed  the gauge symmetry breaking 
by the adiabatic approximation in the ASCC method.
A particular emphasis is put on the gauge symmetry breaking 
for the non-point gauge transformations.
We have discussed the gauge symmetry breaking due to the adiabatic approximation
to $\hat G$ and a possible extension of the ASCC method including
the higher-order operators.
As we have seen in Ref.~\cite{Sato2015} and in this paper,
there are two sources of the gauge symmetry breaking in the ASCC method.
One is the decomposition of the equation of collective submanifold
into the three equations, namely the moving-frame HFB \& QRPA equations, 
depending on the order of $p$.
Example 3 is one example of the gauge symmetry breaking by the decomposition~\cite{Sato2015}.
The other is the truncation of the adiabatic expansion we have 
discussed in details in this paper.

According to the generalized Thouless theorem~\cite{Mang1968}, 
to describe states which are not orthogonal to the vacuum $|\phi(q)\rangle$,
it is sufficient to include only $a^\dagger a^\dagger$ and $aa$ terms
(so-called A-terms) in $\hat G$, and the $a^\dagger a$ terms (so-called
B-terms) are not necessary.
However, when the expansion of $\hat G$ is truncated to the first-order,
$[\tilde N, \hat Q]=0$ is required for the gauge symmetry of the
moving-frame QRPA equations, which implies that $\hat Q$ should contain
B-terms.
In Refs. \cite{Hinohara2007, Hinohara2008, Hinohara2009},
Hinohara et al. required $[\tilde N, \hat Q]=0$ and successfully 
solve the moving-frame HFB \& QRPA equations numerically.
It may be one justification for the requirement of $[\tilde N, \hat Q]=0$
that one can keep the gauge symmetry which exists in the equation of collective
submanifold before the adiabatic expansion. 

In this paper, to conserve the gauge symmetry, we have introduced the higher-order
operators $\hat Q^{(i)}\,\,\,(i>1)$, instead of introducing the B-part
of $\hat Q$ and requiring $[\tilde N, \hat Q]=0$.
It would be interesting to investigate the correspondence between 
the two approaches; one is the approach with the higher-order operators consisting
of only A-terms and the other with only the first-order operator
containing B-terms as well as the A-terms.
This will be investigated in a future publication.


\section*{Acknowledgments}
The author thanks K. Matsuyanagi and N. Hinohara for
their fruitful discussions and comments.

\appendix
\section{Derivation of basic equations}

We derive the basic equations in the cases where $\hat G$ is
expanded up to the second order and up to the third order.
They are derived in a parallel way to Ref. \cite{Matsuo2000}.

We start with the expansion up to the second order,
\begin{equation}
\hat G(q,p,n) =p \hat Q^{(1)}(q)
+n \hat \Theta^{(1)}(q)+\frac{1}{2}p^2 \hat Q^{(2)}(q) +\frac{1}{2}n^2 \hat \Theta^{(2)} (q) +pn \hat X.
\end{equation}
Using the Hadamard lemma \cite{Hall2003},
\begin{align}
e^{X} Y e^{-X} &=e^{\text{ad}_X}Y=Y+[X, Y]+\frac{1}{2}[X,[X,Y]]+ \frac{1}{3!}[X,[X,[X,Y]]]+\cdots ,\label{eq:Hadamard}
\end{align}
we expand the collective Hamiltonian up to the second order
\begin{align}
\mathcal{H}(q,p,n) &=\langle \phi (q,p,\varphi,n) |\hat H |\phi (q,p,\varphi,n) 
\rangle
=\langle \phi(q) | e^{-i\hat  G (q,p,n)} \hat H e^{i\hat  G
 (q,p,n)}|\phi(q)\rangle \notag \\
&= V(q)+\frac{1}{2}B(q)p^2 +\lambda n +\frac{1}{2}D(q)n^2 
\end{align}
with
\begin{align}
 V(q)&=\langle\phi (q)|\hat H |\phi(q)\rangle, \\
 B(q)&=\langle\phi (q)|[\hat H, i\hat  Q^{(2)}] |\phi(q)\rangle 
      -\langle\phi (q)|[[\hat H, \hat  Q^{(1)}],\hat Q^{(1)}] |\phi(q)\rangle, \\
 \lambda(q)&=\langle\phi (q)|[\hat H,i\hat \Theta^{(1)}] |\phi(q)\rangle, \\
 D(q)&=\langle\phi (q)|[\hat H, i\hat  \Theta^{(2)}] |\phi(q)\rangle 
      -\langle\phi (q)|[[\hat H, \hat  \Theta^{(1)}],\hat \Theta^{(1)}] |\phi(q)\rangle. 
\end{align}
Here we have used
\begin{align}
 &\langle \phi(q) | [\hat H, \hat  Q^{(1)}] |\phi(q)\rangle =0, \\
 &\langle \phi(q) |i[\hat H,\hat X]-\frac{1}{2}( [[\hat H, \hat  \Theta^{(1)}],\hat Q^{(1)}]+ [[\hat H, \hat  Q^{(1)}],\hat \Theta^{(1)}]) |\phi(q)\rangle =0. 
\end{align}
Because we are interested in the gauge symmetry, we adopt
the collective Hamiltonian up to $O(n)$,
\begin{equation}
\mathcal{H}(q,p,n) = V(q)+\frac{1}{2}B(q)p^2 +\lambda (q)n . 
\end{equation}
The moving-frame HFB \& QRPA equations are derived from

\noindent
\underline{Eq. of collective submanifold:}
\begin{align}
& \delta \langle \phi(q,p,\varphi, n)|\hat H -
  \frac{\partial \mathcal{H}}{\partial p}\mathring{P}
- \frac{\partial \mathcal{H}}{\partial q}\mathring{Q}
- \frac{\partial \mathcal{H}}{\partial \varphi}\mathring{\Theta}
- \frac{\partial \mathcal{H}}{\partial n}\tilde{N}
|\phi(q,p,\varphi, n) \rangle  &=0, \label{eq:coll.man1} \\
\Longleftrightarrow & \delta \langle \phi(q,p,n)|\hat H -
  \frac{\partial \mathcal{H}}{\partial p}\mathring{P}
- \frac{\partial \mathcal{H}}{\partial q}\mathring{Q}
- \frac{\partial \mathcal{H}}{\partial n}\tilde{N}
|\phi(q,p,n) \rangle  &=0, \label{eq:coll.man2} \\
\Longleftrightarrow & \delta \langle \phi(q)|e^{-i\hat G(q)}\hat H
 e^{i\hat G(q)}-
  \frac{\partial\mathcal{H}}{\partial p}\mathring{P^\prime}
- \frac{\partial\mathcal{H}}{\partial q}\mathring{Q^\prime}
- \frac{\partial\mathcal{H}}{\partial n}\tilde{N^\prime}
|\phi(q) \rangle  &=0 ,\label{eq:coll.man3} 
\end{align}
with 
$\mathring{P^\prime}=e^{-i\hat G}\mathring{P}e^{i\hat G},
 \mathring{Q^\prime}=e^{-i\hat Q}\mathring{Q}e^{i\hat G},$
and 
$\tilde{N^\prime}=e^{-i\hat G}\tilde{N}e^{i\hat G}$.
For the first equivalence, $\partial \mathcal{H}/\partial \varphi=0$ is
used.

The canonical-variable conditions are derived from the canonicity conditions.

\noindent
\underline{Canonicity conditions:}
\begin{align}
 \langle \phi(q,p,\varphi,n)|\mathring{P} |\phi(q,p,\varphi, n) \rangle&
=\langle \phi(q)|\mathring{P^\prime} |\phi(q) \rangle
 =p, \label{eq:canonicity 1 with s=0}\\
   \langle \phi(q,p,\varphi,n)  | \mathring{Q}|\phi(q,p,\varphi, n) \rangle&
=\langle \phi(q)|\mathring{Q^\prime} |\phi(q) \rangle
=0, \\
   \langle \phi(q,p,\varphi,n)|\tilde{N} |\phi(q,p,\varphi, n)
 \rangle&
=\langle \phi(q)|\tilde{N^\prime} |\phi(q) \rangle
 =n, \\
         \langle \phi(q,p,\varphi,n)| \mathring{\Theta}|\phi(q,p,\varphi, n) \rangle&
=\langle \phi(q)|\mathring{\Theta^\prime} |\phi(q) \rangle
 =0, \label{eq:canonicity 4 with s=0} 
\end{align}
with $\mathring{\Theta^\prime}=e^{-i\hat G}\mathring{\Theta}e^{i\hat G}$.
With the use of the general formula (\ref{eq:Hadamard}),
the unitary transformations of the generators are expanded as
\begin{align}
\mathring{P}^\prime
&=i\partial_q - p\partial_q \hat Q^{(1)}-n\partial_q \hat \Theta^{(1)}
 -pn\partial_q \hat X
-\frac{1}{2}p^2\left(i[\partial_q \hat Q^{(1)},\hat Q^{(1)}]+\partial_q \hat Q^{(2)}\right)\notag \\
&  -\frac{1}{2}n^2\left(i[\partial_q \hat \Theta^{(1)},\hat \Theta^{(1)}]
 +\partial_q \hat \Theta^{(2)}\right) -\frac{i}{2} pn \left([\partial_q \hat Q^{(1)},\hat \Theta^{(1)}] 
                      +[\partial_q \hat \Theta^{(1)},\hat Q^{(1)}]
 \right)+\cdots  \notag\\
&=\hat P +i p[\hat P, \hat Q^{(1)}]+i n[\hat P, \hat \Theta^{(1)}]+\frac{1}{2}p^2\left(i[\hat P,\hat Q^{(2)}]-[[\hat P,\hat Q^{(1)}],\hat Q^{(1)}]\right) \notag \\
&+\frac{1}{2}n^2\left(i[\hat P,\hat \Theta^{(2)}]-[[\hat P, \hat\Theta^{(1)}],\hat\Theta^{(1)}] \right) \notag\\
&+ pn \left(i[\hat P,\hat X] -\frac{1}{2}([[\hat P,\hat Q^{(1)}],\hat\Theta^{(1)}] 
                      +[[\hat P, \hat\Theta^{(1)}],\hat Q^{(1)}])
                      \right) +\cdots , \label{eq: mathring P with X} 
\end{align}
\begin{align}
\mathring{Q}^\prime
&=\hat Q^{(1)} +p \hat Q^{(2)} 
+n(\hat X + \frac{i}{2}[    \hat Q^{(1)}, \hat \Theta^{(1)}])\notag \\
&-\frac{i}{4}p^2[ \hat Q^{(1)}, \hat Q^{(2)}]
+\frac{1}{2}n^2\left(\frac{i}{2}[\hat Q^{(1)},\hat \Theta^{(2)}]
 -\frac{1}{3}[[\hat Q^{(1)}, \hat \Theta^{(1)}], \hat \Theta^{(1)}]
+i[\hat X,\hat \Theta^{(1)}] \right)\notag \\
&+\frac{pn}{2}  \left(i[\hat Q^{(2)},\hat \Theta^{(1)}]
 -\frac{1}{3}[[\hat Q^{(1)}, \hat \Theta^{(1)}] ,\hat Q^{(1)}]\right)+
 \cdots , \label{eq: mathring Q with X} \\
\mathring{\Theta}^\prime
&=\hat \Theta^{(1)} +n \hat \Theta^{(2)} +p(\hat X+\frac{i}{2}[\hat\Theta^{(1)},\hat Q^{(1)}])\notag \\
& -\frac{i}{4}n^2[ \hat \Theta^{(1)}, \hat \Theta^{(2)}]
+\frac{1}{2}p^2\left(\frac{i}{2} [\hat\Theta^{(1)}, \hat Q^{(2)}]
 -\frac{1}{3}[[\hat \Theta^{(1)}, \hat Q^{(1)}], \hat Q^{(1)}]
+i[\hat X,\hat Q^{(1)}])  \right)\notag \\
& +\frac{1}{2}pn \left(i[\hat \Theta^{(2)}, \hat Q^{(1)}]
-\frac{1}{3}[[\hat \Theta^{(1)}, \hat Q^{(1)}] ,\hat
 \Theta^{(1)}]\right) \cdots ,  \label{eq:2nd-order mathring Theta} \\
\tilde{N}^\prime&=\tilde N +  ip[ \tilde N, \hat Q^{(1)}] +in [ \tilde N, \hat \Theta^{(1)}] \notag \\
&+ \frac{1}{2}p^2 \left(i[ \tilde N, \hat Q^{(2)}] -[[ \tilde N, \hat Q^{(1)}], \hat Q^{(1)}]\right)
  +\frac{1}{2}n^2 \left(i[ \tilde N, \hat \Theta^{(2)}] -[[ \tilde N, \hat \Theta^{(1)}],\hat \Theta^{(1)}]\right) \notag\\ 
&+pn \left(i[\tilde N,\hat X]-\frac{1}{2}([[\tilde N , \hat Q^{(1)}],\hat
 \Theta^{(1)}]+[[\tilde N , \hat \Theta^{(1)}],\hat Q^{(1)}]) \right)
 +\cdots .\label{eq: mathring N with X} 
\end{align}
By substituting (\ref{eq: mathring P with X})- (\ref{eq: mathring N with
X})
to (\ref{eq:canonicity 1 with s=0})-(\ref{eq:canonicity 4 with s=0}),
we obtain the zeroth- and first-order canonical-variable conditions (\ref{eq: 0th CVC with
 X 1})-(\ref{eq:1st cvc with X 8}).
[One can also define the second-order canonical-variable conditions at
this point. 
However, as shown in Eqs. (\ref{eq:2nd CVC w Q3 1})-(\ref{eq:2nd CVC w Q3 12}), 
the second-order canonical-variable conditions contain
contributions from the third-order operators, which are not taken into
account now.]

By substituting (\ref{eq: mathring P with X})- (\ref{eq: mathring N with
X})
to the equation of collective submanifold, we obtain

\noindent
\underline{ The zeroth-order equation}
\begin{equation}
 \delta \langle \phi(q)|\hat H_M |\phi(q)\rangle =0,
 \label{eq:moving-frame HFB in App. B}
\end{equation}

\noindent
\underline{ The equation of the order of  $p$}
\begin{equation}
 \delta \langle \phi(q)|[\hat H_M , \hat Q^{(1)}] 
-\frac{1}{i}B(q)\hat P -\frac{1}{i}\partial_q V \hat Q^{(2)} |\phi(q)\rangle =0,
 \label{eq:moving-frame QRPA1 in App. B}
\end{equation}

\noindent
\underline{ The equation of the order of  $p^2$}
\begin{align}
 \delta \langle \phi(q)|\frac{1}{2}[[\hat H_M ,\hat Q^{(1)}], \hat Q^{(1)}] 
-B(q) \Delta\hat Q^{(1)} 
-\frac{i}{2}[\hat H-\lambda \tilde N ,\hat Q^{(2)}]  
-\frac{i}{4}\partial_q V[\hat Q^{(1)} ,\hat Q^{(2)}]  
|\phi(q)\rangle =0,  \label{eq:p^2 EoCSM}  
\end{align}
where 
\begin{align}
\hat H_M=\hat H -\lambda \tilde N -\partial_q\hat Q^{(1)},\\
\Delta Q^{(1)}= \partial_q \hat Q^{(1)}+\Gamma (q)\hat Q^{(1)},\\
\Gamma (q)=-\frac{1}{2B(q)}\partial_q B(q).
\end{align}
We take
the first derivative of the zeroth-order equation with respect
to $q$ and obtain
\begin{align}
 \delta \langle \phi(q)|
[\hat H_M, \frac{1}{i}\hat P]
-C(q)\hat Q^{(1)}-\partial_q V \Delta \hat Q^{(1)}-\partial_q \lambda
 \tilde N|\phi(q)\rangle =0,\label{eq:1st der. of mfHFB}
\end{align}
with $C(q)=\partial_q^2 V-\Gamma (q)\partial_qV$.
We eliminate $\Delta \hat Q^{(1)}$ from Eq. (\ref{eq:p^2 EoCSM})  
with use of Eq. (\ref{eq:1st der. of mfHFB}),
which leads to  the moving-frame QRPA equation of $O(p^2)$
\begin{align}
 \delta \langle \phi(q)|
[\hat H -\lambda \tilde N -\partial q V \hat Q^{(1)}, \frac{1}{i}\hat P]
-C(q)\hat Q^{(1)}-\partial_q \lambda \tilde N \notag\\
-\frac{1}{2B}\left\{[[\hat H-\lambda \tilde N, \partial_q V \hat
 Q^{(1)}],\hat Q^{(1)}] -i \partial_q V [\hat H-\lambda \tilde N, \hat
 Q^{(2)}] \right.\notag \\
\left. -\frac{i}{2}(\partial_q V)^2 [\hat Q^{(1)}, \hat Q^{(2)}]\right\}
|\phi(q)\rangle=0. \label{eq:moving-frame QRPA2 in App. B}
\end{align}
Equations (\ref{eq:moving-frame HFB in App. B}), (\ref{eq:moving-frame QRPA1
in App. B}) and (\ref{eq:moving-frame QRPA2 in App. B}) are the
moving-frame HFB \& QRPA equations in the case of the second-order
expansion of $\hat G$.

Then we move to the expansion of $\hat G$ up to the third order
\begin{align}
\hat G(q,p,n) &=p \hat Q^{(1)}(q)+n \hat \Theta^{(1)}(q)
+\frac{1}{2}p^2 \hat Q^{(2)}(q) 
+\frac{1}{2}n^2 \hat \Theta^{(2)} (q) +pn \hat X \notag\\
&+\frac{1}{3!}p^3 \hat Q^{(3)}(q) 
+\frac{1}{3!}n^3 \hat \Theta^{(3)}(q) 
+\frac{1}{2}p^2n \hat O^{(2,1)}(q) 
+\frac{1}{2}pn^2 \hat O^{(1,2)}(q) .
\end{align}
While the third-order operators do not contribute to 
$\mathring{P^\prime}$, $\tilde N^\prime$ and  $e^{-i\hat G}\hat H
e^{i\hat G}$ 
up to the second order, 
they are involved in the second-order terms of
$\mathring {Q^\prime}$ and $\mathring {\Theta^\prime}$ as follows.
\begin{align}
\mathring{Q}^\prime
&=\hat Q^{(1)} +p \hat Q^{(2)} 
+n(\hat X + \frac{i}{2}[    \hat Q^{(1)}, \hat \Theta^{(1)}])
+\frac{1}{2}p^2 \left( \hat Q^{(3)} -\frac{i}{2}[\hat Q^{(1)},\hat Q^{(2)}]\right)\notag\\
&+\frac{1}{2}n^2 \left( \hat O^{(1,2)} +\frac{i}{2}[\hat Q^{(1)},\hat \Theta^{(2)}]
-\frac{1}{3}[[\hat Q^{(1)}, \hat \Theta^{(1)}], \hat \Theta^{(1)}]+ i[\hat X, \Theta^{(1)}]
\right)\notag\\
&+pn \left(  
 \hat O^{(2,1)} +\frac{i}{2}[\hat Q^{(2)},\hat \Theta^{(1)}]
+\frac{1}{3!}[[\hat \Theta^{(1)}, \hat Q^{(1)}], \hat Q^{(1)}]
\right)
 \cdots,  \label{eq:3rd-order mathring Q} 
\end{align}
\begin{align}
\mathring{\Theta}^\prime
&=\hat \Theta^{(1)} +n \hat \Theta^{(2)} +p(\hat X+\frac{i}{2}[\hat\Theta^{(1)},\hat Q^{(1)}])\notag 
+\frac{1}{2}n^2 \left( \hat \Theta^{(3)} -\frac{i}{2}[\hat
 \Theta^{(1)},\hat \Theta^{(2)}]\right)
\notag \\
&+\frac{1}{2}p^2 \left( \hat O^{(2,1)} +\frac{i}{2}[\hat \Theta^{(1)},\hat Q^{(2)}]
-\frac{1}{3}[[\hat \Theta^{(1)}, \hat Q^{(1)}], \hat Q^{(1)}]+ i[\hat X, Q^{(1)}]
\right)\notag \\
&+pn \left(  
 \hat O^{(1,2)} +\frac{i}{2}[\hat \Theta^{(2)},\hat Q^{(1)}]
+\frac{1}{3!}[[\hat Q^{(1)}, \hat \Theta^{(1)}], \hat \Theta^{(1)}]+
\right)
\cdots . \label{eq:3rd-order mathring Theta} 
\end{align}
With (\ref{eq: mathring P with X}), (\ref{eq: mathring N with X}), 
(\ref{eq:3rd-order mathring Q}) and  (\ref{eq:3rd-order mathring Theta}), 
we obtain the second-order canonical-variable conditions 
(\ref{eq:2nd CVC w Q3 1})-(\ref{eq:2nd CVC w Q3 12}). 
The moving-frame HFB \& QRPA equations are also readily derived;
The equations of $O(1)$ and $O(p)$ are unchanged after the inclusion of
the third-order operators. 
The moving-frame QRPA equation of $O(p^2)$ is derived similarly to the
case of the second-order expansion and now given by
\begin{align}
 \delta \langle \phi(q)|
[\hat H -\lambda \tilde N -\partial_q V \hat Q^{(1)}, \frac{1}{i}\hat P]
-C(q)\hat Q^{(1)}-\partial_q \lambda \tilde N \hspace{18em} \notag\\
-\frac{1}{2B}
\partial_q V
\left\{[[\hat H-\lambda \tilde N,  \hat Q^{(1)}],\hat Q^{(1)}] -i [\hat H-\lambda \tilde N, \hat
 Q^{(2)}] 
+\partial_q V( \hat Q^{(3)}-\frac{i}{2} [\hat Q^{(1)}, \hat Q^{(2)}])\right\}
|\phi(q)\rangle =0. \label{eq:moving-frame QRPA2 with Q3 in App.}
\end{align}
Equations (\ref{eq:moving-frame HFB in App. B}), (\ref{eq:moving-frame QRPA1
in App. B}) and (\ref{eq:moving-frame QRPA2 with Q3 in App.}) are the
moving-frame HFB \& QRPA equations in the case of the third-order expansion.

\section{Gauge transformation in the case of the third-order expansion}
In Sect. 4, we discuss the gauge symmetry in Example 1 when $\hat G$ is
taken up to the third order.
In this Appendix, we briefly discuss the gauge symmetry in Examples 2-4. 

\subsection*{Example 2}
In this example, $G=\frac{\alpha}{2}n^2$ generates $\varphi\rightarrow \varphi +\alpha n$.
By transforming the state vector as
\begin{align}
& e^{-i\varphi \tilde N}e^{i\hat G}|\phi(q)\rangle \notag \\
\rightarrow &  e^{-i(\varphi -\alpha n) \tilde N}e^{iG}|\phi(q)\rangle
 \notag \\
= &  e^{-i\varphi \tilde N}\exp\left\{
i\hat G+i\alpha n \tilde N+\frac{1}{2}[i\alpha n \tilde N, i \hat G] \right.\notag\\
&\left. \hspace{4em} +\frac{1}{12}(
[i\alpha n\tilde N,[i\alpha n\tilde N, i\hat G]]+[i\hat G,[i\hat G,i\alpha n\tilde N]]
)
+\cdots
 \right\}|\phi(q)\rangle \notag \\
= &  e^{-i\varphi \tilde N}\exp\left\{
i p\hat Q^{(1)} + in (\Theta^{(1)}+\alpha \tilde N) +\frac{i}{2}p^2\hat
 Q^{(2)} \right.\notag \\
& +\frac{i}{2}n^2( \hat  \Theta^{(2)}+i\alpha[\tilde N, \hat\Theta^{(1)}] )
+ pn(\hat X+\frac{i}{2}\alpha  [\tilde N,  \hat Q^{(1)} ]) \notag\\
&+\frac{i}{6}p^3\hat Q^{(3)}
+\frac{i}{6}n^3\left( \hat \Theta^{(3)} + \frac{3}{2}i\alpha
 [\tilde N, \hat \Theta^{(2)}] -\frac{1}{2}\alpha[\hat
 \Theta^{(1)},[\hat \Theta^{1},\tilde N]] \right) \notag\\
&+\frac{i}{2}p^2n\left(\hat O^{(2,1)} + \frac{i}{2}\alpha [\tilde
 N, \hat Q^{(2)}] 
-\frac{1}{6}\alpha[\hat Q^{(1)},[\hat Q^{1},\tilde N]]\right) \notag\\
&\left.+\frac{i}{2}pn^2\left(
\hat O^{(1,2)} + i\alpha [\tilde
 N, \hat X] 
-\frac{1}{6}\alpha\left([\hat Q^{(1)},[\hat \Theta^{1},\tilde N]]+[\hat \Theta^{(1)},[\hat Q^{1},\tilde N]]\right)
\right)
 \right\}
|\phi(q)\rangle,
\end{align}
we find the following transformation.
\begin{align}
&\hat \Theta^{(1)}\rightarrow \hat \Theta^{(1)}+ \alpha \tilde N, \\
&\hat \Theta^{(2)}\rightarrow \hat \Theta^{(2)}+ i\alpha [\tilde N, \hat
 \Theta^{(1)}], \\
&\hat X \rightarrow \hat X + \frac{i}{2}\alpha  [\tilde N, \hat Q^{(1)}], \\
&\hat \Theta^{(3)} \rightarrow \hat \Theta^{(3)} + \frac{3}{2}i\alpha
 [\tilde N, \hat \Theta^{(2)}] -\frac{1}{2}\alpha[\hat
 \Theta^{(1)},[\hat \Theta^{1},\tilde N]] ,\\
&\hat O^{(2,1)} \rightarrow \hat O^{(2,1)} + \frac{i}{2}\alpha [\tilde
 N, \hat Q^{(2)}] 
-\frac{1}{6}\alpha[\hat Q^{(1)},[\hat Q^{1},\tilde N]],\\
&\hat O^{(1,2)} \rightarrow \hat O^{(1,2)} + i\alpha [\tilde
 N, \hat X] 
-\frac{1}{6}\alpha\left([\hat Q^{(1)},[\hat \Theta^{1},\tilde N]]+[\hat \Theta^{(1)},[\hat Q^{1},\tilde N]]\right).
\end{align}
It is clear that the moving-frame HFB equation and the moving-frame
QRPA equations of $O(p)$ and $O(p^2)$ are invariant under the above
transformation.
The second-order canonical-variable conditions are also gauge invariant.

\subsection*{Example 3}
$G=\alpha qn$ generates $\varphi\rightarrow \varphi
+\alpha q $ and $p\rightarrow p - \alpha n $.
By considering the following transformation,
\begin{align}
& e^{-i\varphi \tilde N}e^{i\hat G}|\phi(q)\rangle \notag \\
\rightarrow &  e^{-i(\varphi -\alpha q) \tilde N}
\exp \left\{ i\left[ (p+\alpha n)\hat Q^{(1)} 
+ \frac{1}{2}(p+\alpha n)^2 \hat Q^{(2)} +n \hat \Theta^{(1)} + \frac{1}{2}n^2 \hat \Theta^{(2)}\right.\right.\notag\\
&\hspace{4em}
+ (p+\alpha n)n\hat X	
+\frac{1}{6}(p+\alpha n)^3\hat Q^{(3)} 
+\frac{1}{2}(p+\alpha n)^2n\hat O^{(2,1)} \notag\\
&\hspace{8em}\left. \left. 
+\frac{1}{2}(p+\alpha n)n^2\hat O^{(1,2)} 
+\frac{1}{6}n^3\hat \Theta^{(3)} 
      \right]\right\}|\phi(q)\rangle  \notag \\
& = e^{-i\varphi\tilde N}e^{-i q(\hat P-\alpha \tilde N)}
\exp \left\{ i\left[ p\hat Q^{(1)} +n (\hat \Theta^{(1)}+\alpha \hat Q^{(1)}) 
+ \frac{1}{2}p^2 \hat Q^{(2)} \right.\right.\notag\\
&+ \frac{1}{2}n^2 (\hat \Theta^{(2)}+2\alpha \hat X)+ pn(\hat X+\alpha
 \hat Q^{(2)})
+\frac{1}{6}p^3\hat Q^{(3)} \notag\\
&\hspace{0em}\left. \left.  
+\frac{1}{2}p^2n\left(\hat O^{(2,1)}+ \alpha \hat Q^{(3)}\right)
+\frac{1}{2}pn^2\left(\hat O^{(1,2)}+ 2\alpha \hat O^{(2,1)} \right)
+\frac{1}{6}n^3\left(\hat \Theta^{(3)}+ 3\alpha \hat O^{(1,2)} \right)
\right]\right\}|_{q=0}|\phi(0)\rangle,  
\end{align}
we find
\begin{align}
& \hat P \rightarrow \hat P - \alpha \tilde N, \\
&\hat \Theta^{(1)}\rightarrow \hat \Theta^{(1)}+ \alpha\hat Q^{(1)}, \\
&\hat \Theta^{(2)}\rightarrow \hat \Theta^{(2)}+ 2\alpha \hat X, \\
&\hat X \rightarrow \hat X + \alpha\hat Q^{(2)},\\
&\hat \Theta^{(3)}\rightarrow \hat \Theta^{(3)}+ 3\alpha \hat O^{(1,2)}, \\
&\hat O^{(2,1)}\rightarrow \hat O^{(2,1)}+ \alpha \hat Q^{(3)}, \\
&\hat O^{(1,2)}\rightarrow \hat O^{(1,2)}+ 2\alpha \hat O^{(2,1)}. 
\end{align}
As in the case of the second-order expansion in Sect. 3, 
the moving-frame HFB equation is gauge invariant, while the moving-frame
QRPA equation of $O(p)$ is not.
The moving-frame QRPA equation of $O(p^2)$ is not gauge invariant
because $[\tilde N, \hat Q^{(1)}] \neq 0$.
The canonical-variable conditions up to the second order are gauge invariant.

\subsection*{Example 4}
$G=\alpha \varphi n$ generates
$\varphi \rightarrow (1+\alpha)\varphi = e^\alpha \varphi$
and $n \rightarrow (1-\alpha)n = e^{-\alpha} n$.
Therefore, by considering 
\begin{align}
& e^{-i\varphi \tilde N}e^{i\hat G}|\phi (q) |\phi (q) \rangle \notag \\
\rightarrow &e^{-i(1-\alpha )\varphi\tilde N } 
\exp\left[ i\hat G(q,p,(1+\alpha) n )\right]|\phi (q) \rangle ,\notag
\end{align}
we find
\begin{align}
&\hat \Theta^{(1)}\rightarrow (1+\alpha )\hat \Theta^{(1)}= e^{\alpha }\Theta^{(1)}, \\
&\hat \Theta^{(2)}\rightarrow (1+2\alpha )\hat \Theta^{(2)} = e^{2\alpha }\Theta^{(2)}, \\
&\hat X \rightarrow (1+\alpha)\hat X=e^{\alpha}\hat X, \\
&\tilde N \rightarrow (1-\alpha )\tilde N= e^{-\alpha }\tilde N, \\
&\hat \Theta^{(3)}\rightarrow (1+3\alpha )\hat \Theta^{(3)} = e^{3\alpha }\Theta^{(3)}, \\
&\hat O^{(2,1)}\rightarrow (1+\alpha )\hat O^{(2,1)} = e^{\alpha }O^{(2,1)}, \\
&\hat O^{(1,2)}\rightarrow (1+2\alpha )\hat O^{(1,2)} = e^{2\alpha }O^{(1,2)}. 
\end{align}
The moving-frame HFB \& QRPA equations are gauge invariant.
It is also clear that the canonical-variable conditions are gauge invariant.



\end{document}